\documentclass[a4paper,superscriptaddress,floatfix,nofootinbib,twocolumn,longbibliography]{revtex4}

\usepackage{bbold}
\usepackage{bbm}
\usepackage[pdftex]{graphicx}
\usepackage{latexsym,amsmath,verbatim,amssymb,txfonts}
\usepackage{color}
\usepackage{rotating}
\usepackage{verbatim}
\usepackage{multirow}
\usepackage[english]{babel}
\usepackage{comment}
\usepackage[normalem]{ulem}
\usepackage{pdfpages}
\usepackage{pgffor}

\begin{document} 

\title{Uncertainty reduction for stochastic processes on complex networks}

\author{Filippo Radicchi}
\affiliation{Center for Complex Networks and Systems Research, School
  of Informatics, Computing, and Engineering, Indiana University, Bloomington,
  Indiana 47408, USA}
\email{filiradi@indiana.edu}

\author{Claudio Castellano}
\affiliation{Istituto dei Sistemi Complessi (ISC-CNR), Via dei Taurini 19, 00185 Roma, Italy}

\begin{abstract}
Many real-world systems are characterized by
stochastic dynamical rules where a complex network
of interactions 
among individual elements probabilistically
determines their state. 
Even with full knowledge
of the network structure and of the stochastic rules, 
the ability to predict system configurations
is generally characterized by large uncertainty. 
Selecting a fraction of the nodes and observing 
their state may help to reduce the uncertainty about the unobserved nodes.
However, choosing these points of observation 
in an optimal way is a highly nontrivial task, depending on the nature 
of the stochastic process and on the structure of the underlying 
interaction pattern. 
In this paper, we introduce a computationally efficient algorithm to 
determine quasi-optimal solutions to the 
problem. 
The method leverages network sparsity to reduce computational complexity
from exponential to almost quadratic, thus allowing the straightforward
application of the method to mid-to-large-size systems. 
Although the method is exact only for equilibrium stochastic processes
defined on trees, it turns out to be
effective also 
for out-of-equilibrium processes
on sparse loopy networks.
\end{abstract}

\maketitle

Stochastic phenomena are studied in any field of science,
including
biology~\cite{bressloff2014stochastic},
ecology~\cite{lande2003stochastic}, physics~\cite{van1995stochastic}, 
neuroscience~\cite{laing2010stochastic}, and finance~\cite{viens2002stochastic}.
In a stochastic system composed of multiple elements,
the states of the elements obey probabilistic rules that
depend on the states of other elements.
Often, 
a sparse network describes
how elements interact
one with the other~\cite{newman2010networks}.
Consider flu spreading for example. 
The epidemics starts from a few initial seeds. 
A person not immunized can contract the disease with a certain 
probability only if in contact with an infected individual. 
At the same time, infected people can spontaneously recover.
The social network underlying the spreading process determines how the
state of every individual depends on the others. 
At any given time, the system is characterized by some uncertainty, 
in the sense that different configurations have 
a non-vanishing probability to appear. 
Such an uncertainty is due to the stochasticity 
of the process, and it is present regardless of the 
knowledge possessed about the probabilistic dynamical rules 
and about the contact pattern.

To reduce uncertainty, one can
observe the state of a sample of elements.
In the example of flu spreading, this means
obtaining full knowledge about the health state of some people. 
With such a knowledge, the prediction 
of the state of unobserved elements becomes less uncertain. 
In particular, the larger the sample, the lower the 
uncertainty, with the limiting case of null uncertainty when 
the entire system is observed.
Resource constraints make complete observation usually
impossible. Is there an efficient way of identifying the best elements to observe
so that the uncertainty for the rest of system is minimized?
The question is answered, from an information-theoretic point of view, by 
the principle of maximum entropy sampling (MES)~\cite{shewry1987maximum}. 
Its rationale is intuitive: to reduce uncertainty about the 
system as much as possible, 
the elements for which joint uncertainty is maximal must be observed.
MES is often used as a solution to problems of experimental 
design~\cite{chaloner1995bayesian}.
An example is the problem of where to place
thermometers in a room to provide the most accurate picture 
of the temperature in the entire room~\cite{krause2008near}.
In special settings, MES can be efficiently approximated or 
achieved exactly with ad-hoc 
algorithms~\cite{ko1995exact, lee2002maximum, guestrin2005near},
These studies have, however, considered very small systems because
the computational complexity of the proposed algorithms grows 
exponentially.
Further, the problem has been studied only in regular
topologies, such as lattices or fully connected networks.
The present paper considers the MES 
problem when the interaction pattern is given by
a large complex network.
In this case,
the sparsity of the topology can be
leveraged to make the application of MES
feasible in rather large systems.

To avoid any potential confusion, we stress 
that our goal is the selection of 
a fraction of observed nodes in order to minimize
the uncertainty on the stochastic variables 
associated with unobserved nodes.
This is distinct from the problem of optimally sampling a 
network to reduce uncertainty on its
unknown properties (e.g., degree distribution, 
diameter, size)~\cite{Leskovec2006, thompson2013dynamic}.
Also, our problem is different from
active learning in 
networks~\cite{bilgic2010active,moore2011active, peel2014active}, 
where the goal is to infer a model  
able to predict the value of the unobserved variables in
a specific configuration of the system. 
In our case, we do not infer parameter models.
Further, we are not interested in making predictions
about a specific configuration. Instead, for all possible
configurations that the system may exhibit, we want to identify
what nodes we need to observe in order to 
minimize our uncertainty about such configurations.
In this respect, our problem is similar to the one studied in
Ref.~\cite{7982740}, with the difference 
that we deal with stochastic rather than deterministic systems.

We consider a dynamical stochastic process defined on a 
graph $\varmathbb{G}$, composed of $N$ nodes. 
Every node $i \in \varmathbb{G}$ is characterized by a state
variable $x_i$ that can assume $K$ distinct values;
$\mathbf{x} = (x_1, x_2, \ldots, x_N)$
corresponds to a specific microscopic configuration of the system.
We assume that the process is Markovian 
and that the change of the state of a 
single node is determined only by local 
interactions with the nodes directly connected
to it.  Hence, the graph $\varmathbb{G}$ 
fully determines how microscopic configurations 
are related one to the other.
Let us indicate with $p(\mathbf{x})$ the stationary
probability distribution associated to each of the $K^N$ possible
microscopic configurations that the system can assume.
Despite full knowledge of the  
graph structure and of the stochastic process,
we are still left with potentially large uncertainty 
quantified by the information-theoretic joint entropy
\begin{equation}
\mathcal{H}(\varmathbb{G}) = 
- \sum_{\mathbf{x}} p(\mathbf{x}) \, \log_2
[p(\mathbf{x})]
\; . 
\label{eq:entropy}
\end{equation}

Suppose we can observe a subset of nodes $\varmathbb{O}
\subseteq \varmathbb{G}$. Observing these nodes removes any
uncertainty on their state, and thus conditions the
joint probability distribution of the unobserved part of the graph,
$\varmathbb{G}\setminus \varmathbb{O}$, to the state of the observed nodes
$\varmathbb{O}$, namely $p(x_{u_1}, \ldots, x_{u_{N-O}} | x_{o_1},
\ldots, x_{o_{O}}) = p(\mathbf{x}_{\varmathbb{G}\setminus
  \varmathbb{O}}| \mathbf{x}_{\varmathbb{O}})$, 
where $u_1, \ldots, u_{N-O} \in
\varmathbb{G}\setminus \varmathbb{O}$, $o_1, \ldots, o_{O} \in \varmathbb{O}$,
and we defined
$\mathbf{x}_{\varmathbb{G}\setminus \varmathbb{O}} = (x_{u_1}, \ldots, x_{u_{N-O}})$ 
and $\mathbf{x}_{\varmathbb{O}}  = (x_{o_1}, \ldots, x_{o_{O}})$.
For a particular choice of the set $\varmathbb{O}$, 
the uncertainty about the rest of the system is quantified by the 
conditional entropy
\begin{equation}
\mathcal{H}(\varmathbb{G}\setminus \varmathbb{O} | \varmathbb{O}) = - 
\sum_{\mathbf{x}_{
  \varmathbb{O}} } p(\mathbf{x}_{\varmathbb{O}}) \;   
\sum_{\mathbf{x}_{\varmathbb{G}\setminus
  \varmathbb{O}} } p(\mathbf{x}_{\varmathbb{G}\setminus
  \varmathbb{O}}| \mathbf{x}_{\varmathbb{O}}) \, \log_2 [p(\mathbf{x}_{\varmathbb{G}\setminus
  \varmathbb{O}}| \mathbf{x}_{\varmathbb{O}})]
\; . 
\label{eq:cond_entropy}
\end{equation}
For $\varmathbb{O} = \emptyset$,  Eq.~(\ref{eq:cond_entropy})
is identical to Eq.~(\ref{eq:entropy}). For $\varmathbb{O} =
\varmathbb{G}$, we have instead $\mathcal{H}(\varmathbb{G}\setminus
\varmathbb{O}|\varmathbb{O}) = \mathcal{H}(\emptyset) = 0$.

We look for the optimal selection of a number $O$ of nodes 
such that their observation minimizes the conditional entropy 
of Eq.~(\ref{eq:cond_entropy}). 
In particular, since $\mathcal{H}(\varmathbb{G}\setminus \varmathbb{O} |
\varmathbb{O}) = \mathcal{H}(\varmathbb{G} )
- \mathcal{H}(\varmathbb{O})$, the minimization of
Eq.~(\ref{eq:cond_entropy}) is equivalent to finding the group of
nodes $\varmathbb{O}^*$ having maximum joint entropy, i.e.,
\begin{equation}
\varmathbb{O}^* = \arg \, \max_{\varmathbb{O}} \, \mathcal{H}(\varmathbb{O}) 
\; ,
\label{eq:max_entropy}
\end{equation}
where $\mathcal{H}(\varmathbb{O})  = -
\sum_{\mathbf{x}_{\varmathbb{O}}} \, p(\mathbf{x}_{\varmathbb{O}}) \,
\log_2 [p(\mathbf{x}_{\varmathbb{O}})]$. 
The maximization is performed over all sets $\varmathbb{O}$ of fixed 
size $O$. This principle is known as MES, 
and the associated problem is $NP$-hard~\cite{shewry1987maximum}. 
The exact solution of this optimization requires the consideration
of all possible choices of the set $\varmathbb{O}$, and for each of 
them the computation of the associated joint entropy. 
The computational complexity of both operations 
scales exponentially with $O$.

A quasi-optimal solution can be obtained
at a reduced computational cost, exploiting the sub-modularity
of the entropy function~\cite{krause2012submodular}.
Such a property allows us to implement a greedy strategy, where the set 
of observed nodes is built sequentially, leading to 
a solution provably close to the optimum~\cite{Nemhauser1978}.
The greedy strategy provides a solution corresponding to a value
of the function to be optimized that is at least $(1 - 1/e) = 0.63\ldots$ 
times the value of the global maximum~\cite{Nemhauser1978}.
In the present context, the greedy algorithm 
consists in sequentially adding, to the set of observed nodes, 
the node with maximal entropy conditioned to the set of variables 
already observed. 
More specifically, the algorithm starts at stage $t=0$ with an empty
set, $\varmathbb{O}_{t=0} = \emptyset$.  
The $t$-th point of observation, namely $o_t$, is chosen, 
among the nodes not yet part of the observed set 
$\varmathbb{O}_{t-1} = \{o_1, o_2, \ldots, o_{t-1}\}$,
according to the rule
\begin{equation}
o_t = \arg \, \max_{i \notin \varmathbb{O}_{t-1}} 
\mathcal{H}(i|o_1, \ldots, o_{t-1}) \; .
\label{eq:greedy_max_entropy}
\end{equation}
The algorithm can be run up to arbitrary values $1 \leq t \leq N$.
This procedure addresses the issue of the extensive search over all
possible groups of nodes. However, at every stage $t$,  
the computation of each of the $N-(t-1)$ conditional entropies 
in Eq.~(\ref{eq:greedy_max_entropy}) still requires a 
number of operations scaling as $K^t$. 
This makes the algorithm usable only for constructing 
very small sets of observed nodes.

To make the greedy algorithm applicable to large sets, 
one must introduce approximations to reduce the computational complexity 
of the calculation of $\mathcal{H}$ in Eq.~(\ref{eq:greedy_max_entropy}).
The simplest and most popular ansatz
in the study of processes on networks is the so-called
individual-based mean-field (IBMF) approximation~\cite{pastor2015epidemic}, 
according to which the joint distribution
$p(\mathbf{x})$ is seen as the product of the marginal probabilities
of the individual nodes, i.e.,
$p(x_1, x_2, \ldots, x_N) = p(x_1) p(x_2) \cdots p(x_N)$. 
The approximation
allows us to write the entropy of any set $\varmathbb{O}$ of
variables as
\begin{equation}
\mathcal{H}_{ind}(\varmathbb{O}) = \sum_{o \in \varmathbb{O}}
\mathcal{H}(o) \; ,
\label{eq:entropy_ind}
\end{equation}
where $\mathcal{H}(o)$ is the unconditional entropy of the node $o$.
Under this approximation, the optimal set
  $\varmathbb{O}_{ind}$ corresponds to the $O$
  nodes with maximal unconditional entropy.

In this paper, we propose a refined approximation,
much less drastic than the IBMF approach, based on two assumptions. 
First, we assume that
the graph $\varmathbb{G}$ fully determines 
dependencies among variables.  
If the pair of
nodes $i$ and $j$ is connected by an edge, then 
the variables $x_i$ and $x_j$ are directly dependent one on the other. 
Otherwise, the variables still depend one on the other 
but only through at least another variable in the system.
This seems a reasonable way of improving the 
IBMF approximation, as we expect that the most 
important dependencies are present among pairs of variables with a 
direct interaction. This assumption is exact for equilibrium
configurations of processes with rates depending on the states
of direct neighbors and satisfying detailed 
balance~\cite{Moussouris1974}.
Second, we assume that the graph 
$\varmathbb{G}$ is a tree. 
Both assumptions are used in our proposed algorithm, that
allows us to efficiently 
compute the entropy, namely $\mathcal{H}_{pair}(\varmathbb{O})$,
for an arbitrary subset of variables $\varmathbb{O}$ [and, as a
consequence, 
the conditional entropies 
in Eq.~(\ref{eq:greedy_max_entropy})].
If the set $\varmathbb{O}$  coincides with the entire graph, then the
algorithm is equivalent to the one used to compute 
the Bethe free-entropy on trees~\cite{mezard2009information}.
The algorithm works sequentially, in the sense that
the function $\mathcal{H}_{pair}(\varmathbb{O})$ is computed by 
iteratively adding single nodes to the set $\varmathbb{O}$. 
This allows us to use the algorithm directly in the
greedy maximization of Eq.~(\ref{eq:greedy_max_entropy}).

Properties of the entropy function alone allow us to write the inequality
\begin{equation}
\mathcal{H}(\varmathbb{O}) \leq \mathcal{H}_{pair}(\varmathbb{O}) \leq
\mathcal{H}_{ind}(\varmathbb{O})
\label{eq:entropy_inequality}
\end{equation}
for any $\varmathbb{O} \subseteq \varmathbb{G}$.
Essentially, our approximation always leads to an upper-bound
of the true entropy function that is tighter than the one predicted
using the standard IBMF approximation.
The approximation is exact, i.e., 
$\mathcal{H}(\varmathbb{O}) \equiv \mathcal{H}_{pair}(\varmathbb{O})$, 
only in the case of equilibrium distributions of systems satisfying 
detailed balance on a tree.

Suppose we are at stage $t$ of the algorithm. We need to
compute the conditional entropy $\mathcal{H}_{pair}(o_{t
}|o_1, \ldots, o_{t-1})$ for the next node $o_t$ that we are adding to
the set. Thanks to Bayes rule, we can write
\begin{equation}
\begin{array}{ll}
\mathcal{H}_{pair}(o_t|o_1, \ldots, o_{t-1}) = & \mathcal{H}(o_t) + \mathcal{H}_{pair}(o_1, \ldots,
o_{t-1}|o_t) 
\\
& - \mathcal{H}_{pair}(o_1, \ldots, o_{t-1}) 
\end{array}
\; .
\label{eq:alg1}
\end{equation}
Under our two main assumptions, the second term can be 
written~(see SM),
as the sum of pairwise conditional entropies, one per observed node
\begin{equation}
\mathcal{H}_{pair}(o_1,\ldots,o_{t-1}|o_t) = \sum_{j=1}^{t-1} \mathcal{H}(o_j|s_{o_j}).
\label{eqtree}
\end{equation}
Every observed node corresponds to a term in the sum, given by the entropy
associated with that observed node, $o_j$, conditioned to another
node $s_{o_j} \in (\varmathbb{O}_{t} \setminus \{o_j\})$. 
Such a node $s_{o_j}$ is either the first observed node encountered along 
the unique  path connecting $o_j$ to $o_t$, or node $o_t$ itself.
Finally, thanks to the chain rule, the rightmost term in 
Eq.~(\ref{eq:alg1}) can be expressed in terms of quantities
computed at previous stages
\begin{equation}
\begin{array}{ll}
\mathcal{H}_{pair}(o_1,\ldots,
o_{t-1}) = &
\mathcal{H}_{pair}(o_{t-1} |o_1, \ldots,
o_{t-2}) +  
\\
& \mathcal{H}_{pair}(o_1, \ldots,
o_{t-2})   
\end{array}
\; .
\label{eq:alg2}
\end{equation}

If the graph is not a tree, the algorithm 
is still applicable as long as the structure is sufficiently
treelike. Many real-world networks
satisfy this condition~\cite{newman2010networks}, and, very often,
treelike approximations are effective even if the graphs
are not treelike~\cite{melnik2011unreasonable}. 
In our proposal, if the graph contains loops, 
$\mathcal{H}_{pair}(o_1, \ldots, o_{t-1}|o_t)$ 
is computed under the tree assumption by generating a spanning
tree rooted in $o_t$, and using again Eq.~(\ref{eqtree}). 
This provides us with an upper-bound of the true entropy, 
since neglecting dependencies necessarily leads 
to an entropy larger than its true value.
The rooted tree can be generated arbitrarily.
However, to keep the upper-bound as tight as possible, we use a 
Djikstra-like algorithm suitably modified for this context~(see SM). 
Results presented here are based on this choice.

The algorithm requires prior knowledge of the unconditional entropy of 
individual nodes, and of the pairwise entropy among pairs of nodes.
From a computational point of view, the running time scales as $N^3$
in the worst-case scenario~(see SM). However,
some computational tricks allow for a great reduction of  the
complexity of the MES algorithm~\cite{leskovec2007cost, krause2012submodular}, 
which effectively scales as $N^2 \log(N)$~(see SM).
This makes the algorithm easily applicable even to relatively large systems. 

To validate the algorithm, we consider four different
processes: i) the Ising model~\cite{RevModPhys.80.1275}; 
ii) the Modified version of
the Susceptible-Infected-Susceptible (MSIS) 
model as proposed in
Ref.~\cite{gleeson2013binary}; iii) the
standard version of the Susceptible-Infected-Susceptible (SIS) 
model~\cite{pastor2015epidemic}; 
iv) the Independent 
Cascade (IC) model~\cite{kempe2003maximizing}. 
The first two models satisfy detailed balance. 
The standard versions of the SIS and IC models are instead
prototypical examples of out-of-equilibrium processes 
that don't satify detailed balance. 
We analyze the behavior of all models for different 
parameter values and on different network substrates, 
including synthetic graphs and real-world networks. 
Results and details of our systematic analysis are reported in the SM.

\begin{figure}[!htb]
\begin{center}
\includegraphics[width=0.45\textwidth]{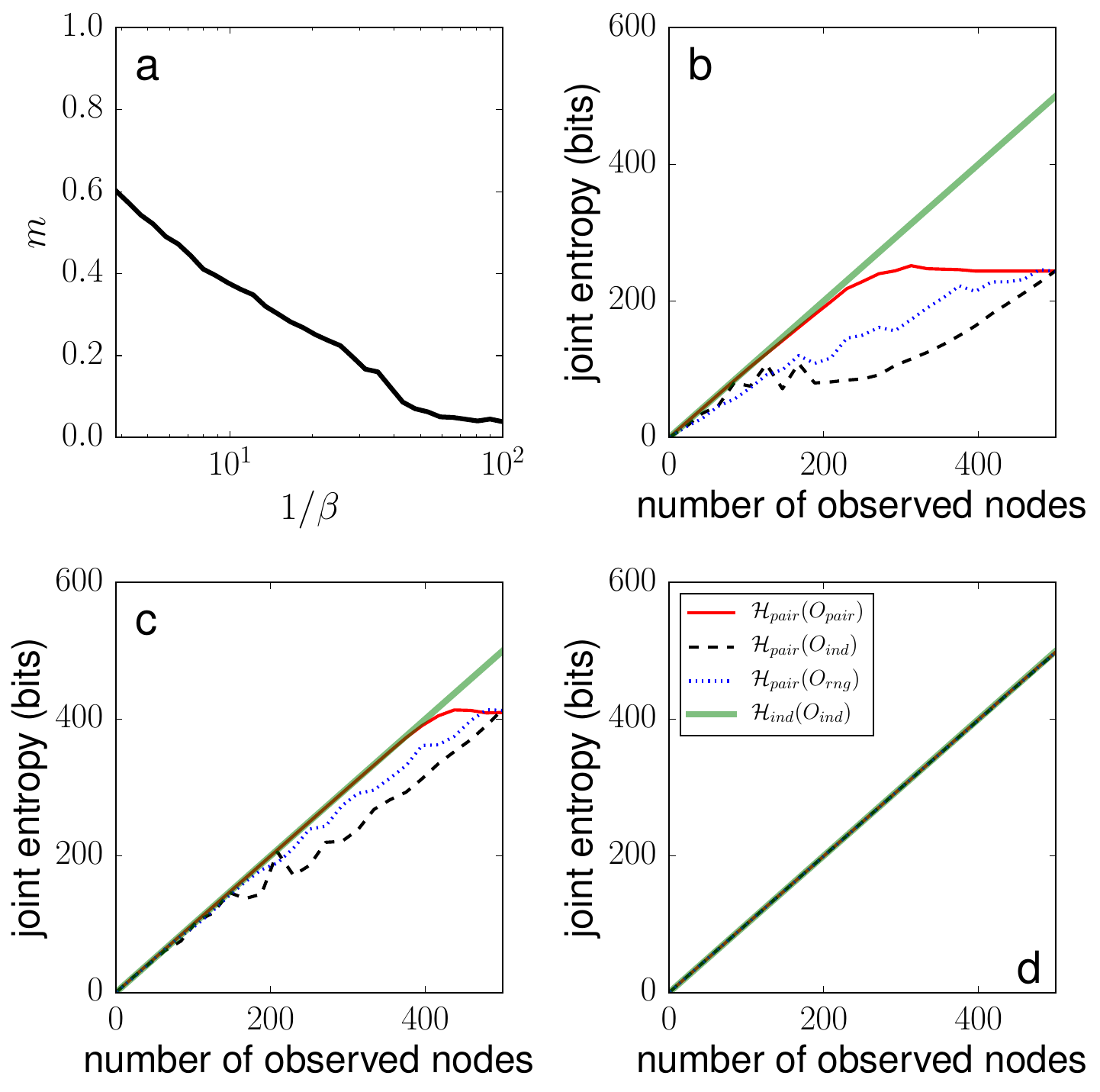}
\end{center}
\caption{Ising model
on the US air transportation network.
a) Magnetization $m$ as a function of the temperature $1/\beta$.
b)  Joint entropy of the observed set as a function of the set size. 
Here $1/\beta = 1$. Sampling techniques considered are:
i) MES (red full line); ii) MES under the IBMF
approximation  
(black dashed line);
iii) random sampling, i.e., nodes are added to the
observed set in random order (blue dotted line). For all sampling
techniques, joint entropy is measured using the novel approximation.
We plot also the joint entropy according to the IBMF approximation 
for the set $\varmathbb{O}_{ind}$ (thick
green line). c) Same as in panel b, but for $1/\beta = 15$. d) Same as
in panels b, but for $1/\beta = 50$.}
\label{fig:ising}
\end{figure}

In Fig.~\ref{fig:ising} we show results
 for the Ising model applied to 
the US air transportation network (size $N=500$)
originally considered in ~\cite{colizza2007reaction}.
The network contains loops, so that our approximation is not exact.
We sample configurations reached by the system after a 
sufficiently long number of iterations of the 
Metropolis algorithm with fixed value of the 
temperature $1/\beta$ and 
external magnetic field $h=1/N$. 
Every realization is obtained after
$1,000 \, N$ total spin flips.
The phase diagram of the system is presented in 
Fig.~\ref{fig:ising}a, 
showing the typical transition from ordered to disordered configurations
as the temperature is increased.
We first analyze statistical properties
of microscopic configurations obtained at
$1/\beta = 1$ in Fig.~\ref{fig:ising}b. 
To estimate the unconditional entropy $\mathcal{H}(i)$ of a 
generic node $i$, and the pairwise conditional 
entropy $\mathcal{H}(j|i)$ of a generic
pair $(i,j)$, we rely on $T = 1,000$ sampled configurations.
In addition to $\varmathbb{O}_{pair}$and  $\varmathbb{O}_{ind}$, 
we consider also
the set of observed nodes $\varmathbb{O}_{rng}$, built 
by adding nodes in random order.
As the figure clearly shows, our approximation 
generates noticeable improvements
with respect to the the IBMF approximation
in the computation of the entropy of subsets of the system.
This is apparent from the large value of the difference 
$\mathcal{H}_{ind}(\varmathbb{O}_{ind})-
\mathcal{H}_{pair}(\varmathbb{O}_{ind})$.
Concerning different sampling strategies, we
also see a significant benefit from using our proposed technique
over the naive version of MES. $\mathcal{H}_{pair}(\varmathbb{O}_{pair})$ grows
much quicker than $\mathcal{H}_{pair}(\varmathbb{O}_{ind})$, and
saturates at the maximum value after about $200$ nodes are observed.
This is an indication that the entire uncertainty of the system can be
explained by looking at a fraction of the nodes in the network only.
On the contrary, $\mathcal{H}_{pair}(\varmathbb{O}_{ind})$ behaves very
similarly to, if not worse than, $\mathcal{H}_{pair}(\varmathbb{O}_{rng})$ testifying that
the naive MES strategy is not effective in this specific
system. As the temperature increases
(Fig.~\ref{fig:ising}c), 
the advantage of using our new approximation in place of the
IBMF approximation becomes less
apparent. At the same time, the benefit of using 
our greedy MES strategy compared to the naive version 
becomes less evident. For very large temperatures, 
all curves become identical (Fig.~\ref{fig:ising}d).

\begin{figure}[!htb]
\begin{center}
\includegraphics[width=0.45\textwidth]{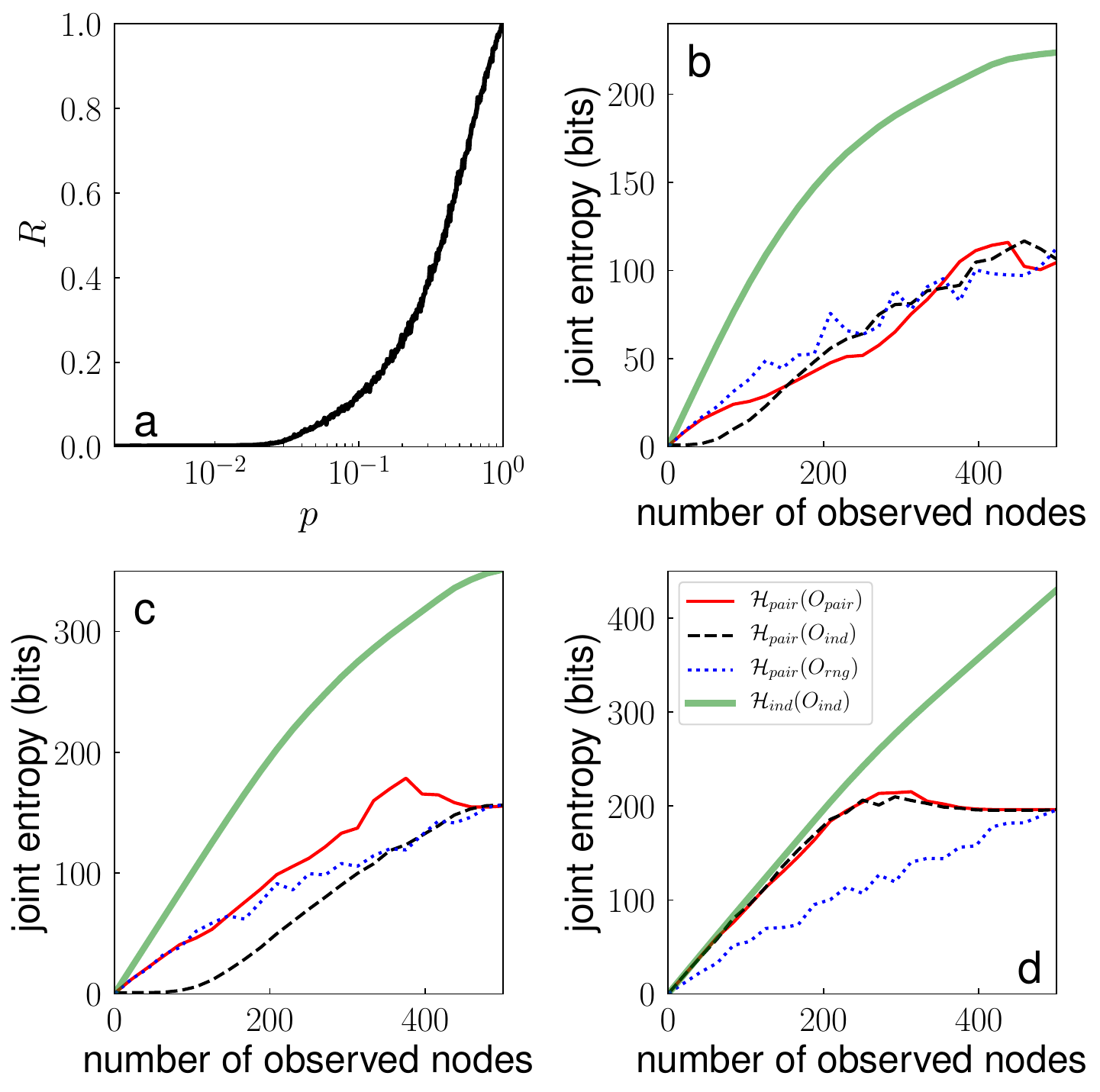}
\end{center}
\caption{
Independent cascade model
on the US air transportation network.
a) Relative size of the outbreak $R$ as a function 
of the infection probability $p$. Panels b, c, and refer respectively
to $p=0.1$, $p=0.2$, and $p=0.5$.
The description of the curves in these panels is identical to
one of the curves appearing in
Fig.~\ref{fig:ising}. 
}
\label{fig:icm}
\end{figure}

In Fig.~\ref{fig:icm}, we study the IC model applied to the same real-world
network. We focus on microscopic
configurations corresponding to the final stage of the dynamics, where
nodes are either in the susceptible or recovered state. The initial
condition of the dynamics is given by all nodes in the susceptible
state, except for a single randomly chosen seed in the infected state.
Infections propagate along each active edge with probability $p$. For every
value of $p$, we consider $T=100,000$ sampled configurations.
In the IC model on a loopy graph, both assumptions
at the basis of our approach are violated. 
Nonetheless, the results reveal that our approximation
represents a significant improvement over the basic
IBMF approximation.
First, we are able to provide estimates of the entropy of the system
that are radically smaller, showing that pairwise
correlations
among variables are particularly significant in the system.
Second, we are able to construct sets of observed nodes
that are more representative for system uncertainty than those
obtained by using the other sampling strategies.

To further strengthen our message, 
in the SM we include a comparison of the
 performance between our greedy algorithm for
MES and other selection strategies: (i)
degree centrality sampling, where nodes are added in decreasing
(increasing) order based on their degree; (ii) closeness centrality
sampling, the same as (i) but with node ranking
based on closeness centrality.  These strategies are chosen
to test the performance of centrality-based metrics 
that rely on topological properties only. 
The most significant difference between them
is that degree is a local metric, whereas closeness is global. 
We find that topological heuristics are not always reliable 
sampling strategies, and that their effectiveness is
seriously affected by the underlying network structure and/or the
parameter values of the stochastic models.
Further in the SM, we study analytically the behavior of the
IC model in star networks and show that the choice
of the best nodes to observe is highly sensitive not only to 
the parameter of the model, but also to the initial configuration 
of the stochastic dynamical process.

In summary, we introduce an algorithm to approximate the conditional 
entropy of a sample of nodes in a complex network.
The algorithm relies on the sparsity of the graph to simplify  
computations otherwise unfeasible. 
Although the algorithm allows us to compute the conditional entropy of 
arbitrary node sets, it finds a particularly interesting application 
in the greedy approximation of the so-called MES principle.
This principle corresponds to the optimal reduction of uncertainty of
a stochastic process taking place on a network. 
Combining our algorithm with machine learning methods
to create active supervised learning approaches is a potentially 
interesting direction for future investigation. 
Other extensions worth of consideration are also the generalization of 
our algorithm to devise computationally feasible selection strategies 
based on other information-theoretic principles, as for example the 
maximization of the mutual information rather than entropy.

\acknowledgements{
  The authors thank A. Puglisi for useful discussions.
FR acknowledges support from the National Science Foundation (Grant
CMMI-1552487), and from the US Army Research Office
(W911NF-16-1-0104).}

\bibliography{bibliography}

\newpage



\section*{Supplementary material}

**SUPPLEMENTAL MATERIAL**

**Reduction of joint conditional entropy in Eq. (7) to a sum of pairwise conditional entropies**

The second term on the r.h.s. of Eq. (7) of the main text can be greatly simplified when the interaction pattern is a tree. We show this in detail by considering the structure depicted in Fig. SM1. In that case the second term of Eq. (7) is $\mathcal{H}_{pair}(o_j, o_w, o_s, o_r|i)$.

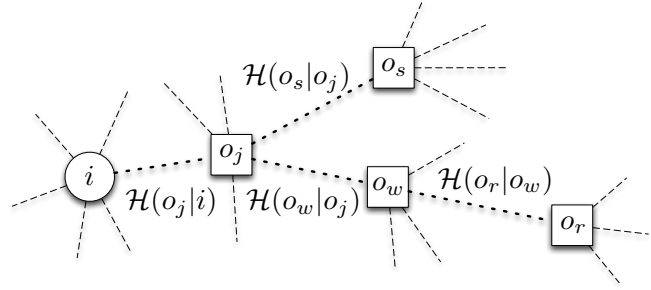

Figure SM1. In a tree, the joint entropy of the observed nodes $o_j$, $o_s$, $o_w$, and $o_r$ conditioned to node $i$ can be broken in the sum of pairwise conditional entropies. Each observed node is responsible for one of these contributions, in the sense that the variable associated with it is conditioned to a variable associated with another node. The node that acts on the conditional part of each pairwise entropy is given either by the first observed node encountered along the path towards $i$ or by node $i$ itself. In the illustration above, the path connecting $o_r$ to $i$ passes through node $o_w$, giving rise to the contribution $\mathcal{H}(o_r|o_w)$. The paths connecting $o_w$ and $o_s$ to $i$ pass both through node $o_j$, giving rise to the contributions $\mathcal{H}(o_s|o_j)$ and $\mathcal{H}(o_w|o_j)$, respectively. There are instead no observed nodes along the path between node $o_j$ and node $i$, thus the contribution of node $o_j$ to the joint entropy is $\mathcal{H}(o_j|i)$.

By using the chain rule it can be written as

$$\mathcal{H}_{pair}(o_j, o_w, o_s, o_r|i) = \mathcal{H}(o_j|i) + \mathcal{H}_{pair}(o_w, o_s, o_r|o_j, i) = \mathcal{H}(o_j|i) + \mathcal{H}_{pair}(o_w, o_s, o_r|o_j) \tag{SM1}$$

where the last step takes into account the tree topology. Again the consideration of the topology indicates that of $o_s$, conditioned to the value of $o_j$ is independent from the values of $o_w$ and $o_r$ conditioned to $o_j$ (conditional independence). Hence

$$\mathcal{H}_{pair}(o_j, o_w, o_s, o_r|i) = \mathcal{H}(o_j|i) + \mathcal{H}(o_s|o_j) + \mathcal{H}_{pair}(o_w, o_r|o_j) \tag{SM2}$$

The application of the chain rule on the last term leads to

$$\mathcal{H}_{pair}(o_j, o_w, o_s, o_r|i) = \mathcal{H}(o_j|i) + \mathcal{H}(o_s|o_j) + \mathcal{H}(o_w|o_j) + \mathcal{H}(o_r|o_w) \tag{SM3}$$

The total conditional entropy is thus reduced to the sum of pairwise conditional entropies, one for each observed node.

The argument can be readily generalized to any tree, by multiple applications of the chain rule and of conditional independence, leading to

$$\mathcal{H}_{pair}(o_1, \ldots, o_{t-1}|i) = \sum_{j=1}^{t-1} \mathcal{H}(o_j|s_{o_j}). \tag{SM4}$$

In the sum, the variable associated with each observed node $o_j$ is conditioned to a variable associated with another node $s_{o_j} \in \{i\} \cup (\mathbb{O}_{t-1} \setminus \{o_j\})$. The node $s_{o_j}$ that acts on the conditional part of each pairwise entropy is given either by the first observed node encountered along the path towards $i$ or by node $i$ itself. Notice that dashed lines in Fig. SM1 need not to be direct connections: other unobserved nodes may lie between observed ones.

**Computational Complexity of the Algorithm**

Our algorithm requires us to have prior knowledge of the unconditional entropy of every single node in the graph. The computation of these quantities scales as $KN$, where $N$ is the number of nodes and $K$ is the number of possible states per node. The algorithm also requires prior knowledge of the pairwise conditional entropy among all pairs of nodes, whose computation requires a time scaling as $K^2 N^2$. At stage $t$ of the algorithm, we need to find the node $i$ that maximizes the entropy

$\mathcal{H}(i|o_1, \ldots, o_{t-1})$. This means that we have to compute the function for every unobserved node. We thus have to walk back on the tree from any already observed node towards the node whose entropy is being computed, visiting all edges $M$ of the tree, with an algorithm that scales as $M$. As we need to perform such an operation for all nodes that are still in the set of unobserved nodes, the computational complexity of estimating the conditional entropy of all unobserved nodes is $N\,M$. These operations must be repeated at any stage of the greedy algorithm adding a factor $N$ to the computational complexity. In conclusion, the computational complexity of the entire algorithm is $N^2\,M \sim N^3$. However, several considerations and computational techniques may be used to speed up the algorithm. First, knowledge of the pairwise entropy is not required for all pairs of nodes. *A priori*, we don't know which pairwise entropies will be used by the algorithm, but we can compute them on-the-fly when they are needed. Second, we know that the entropy of individual nodes conditioned to the set of observed nodes can only decrease during the algorithm, i.e., $\mathcal{H}(i|o_1, \ldots, o_{t-2}) \geq \mathcal{H}(i|o_1, \ldots, o_{t-2}, o_{t-1})$. We can therefore use a lazy algorithm that computes Eq. (7) of the main text for an unobserved node only if needed [1, 2]. To keep track of the ranking of unobserved nodes in a computationally cheap way, we can make use of a standard queue algorithm. Using these tricks, the effective computational complexity of the algorithm is greatly reduced (Fig. SM2). From the numerical analysis, we see that the number of updates required by the lazy implementation of the algorithm at each stage scales logarithmically with the system size.

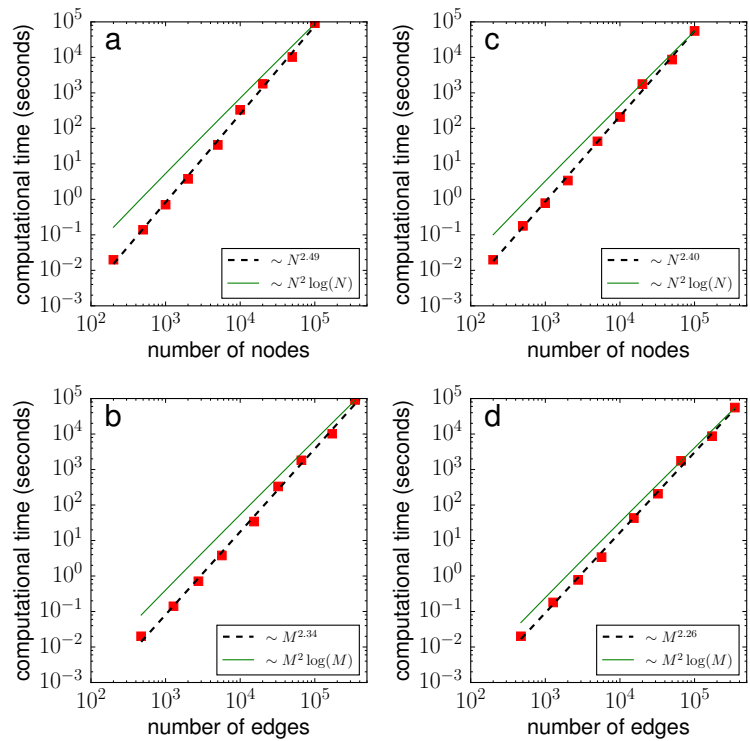

Figure SM2. Computational complexity of the maximum entropy sampling algorithm. We measure the total computational time $C$ required by the algorithm to generate sets of observed nodes from size 1 to $N$. Computations relative to marginal and conditional entropies are not included in our measure. Computations are performed on *Intel(R) Xeon(R) CPU E5-2690 v4 @ 2.60GHz* CPUs. As a representative case we consider here the independent cascade model applied to scale-free networks. The degree distribution of the networks is $P(k) \sim k^{-5/2}$ for $k \in [3, \sqrt{N}]$, and $P(k) = 0$. We further consider different values of the occupation probability $p$. a) Computational time as a function of the network size for $p = 0.1$. The dashed black line is a power-law fit $C \sim N^\beta$, with $\beta = 2.49$. The full green line corresponds instead to a fit $C \sim N^2 \log(N)$. b) Same as in panel a but showing the dependence of the computational time from the total number edges in the network. c) Same as in panel a, but for $p = 0.5$. d) Same as in panel c, but showing the dependence of the computational time from the total number edges in the network.

## Numerical advantages of the algorithm

It is worth to remark the great advantages brought by our approach compared with the naive method [Eq. (2) of the main text] to compute joint conditional entropies when a system is studied with numerical simulations (or experimental observations). This may be the case of most situations. If the size of the set of observed nodes is $O$, the maximum value of the entropy is $O \log_2 K$. This corresponds to the case in which each of the $K^O$ configurations has exactly the same probability to occur. If the number of simulations is $T$, the maximum value of the entropy that can be measured in the naive approach is $\log_2 T$. For large values of $O$, the number of simulations will be unavoidably $T \ll K^O$, thus leading to systematically biased estimates. With our method instead, the maximum value of the pairwise conditional entropy between a pair of nodes is $\log_2 T$. The total entropy of the set of observed nodes will be given by $O$ terms of this type, leading to a maximum possible value equal to $O \log_2 T$. A number of simulations $T \gg K$, computationally feasible in most situations, is then sufficient to avoid numerical problems.

## A Dijkstra-like algorithm to determine the tree with minimal entropy rooted in a specific node

The following provides a description of the algorithm used to determine the tree with minimal entropy rooted in a specific node. The algorithm is a variant of the well-known Dijkstra's algorithm. We used this algorithm whenever we wanted to provide an estimate of the conditional entropy $\mathcal{H}_{pair}(o_1, \ldots, o_{t-1}|i)$ in loopy networks. Here, $i$ is the index of node that acts as a root of the tree, $o_1, o_2, \ldots, o_{t-1}$ are instead the indices of the nodes already observed. We assume that these variables are given, as well as pairwise conditional entropies among pairs of nodes, and the entire topology of the network.

Define $\vec{v}$, with $v_n = -1$ for all $n = 1, \ldots, N$. This vector serves to account for the fact that a node has been already visited. Define $\vec{s}$, with $s_n = -1$ for all $n = 1, \ldots, N$, except for $s_i = i$. This vector serves to keep track of the pairwise conditional entropies that enter in the sum. Define $\vec{o}$, where $o_n = -1$ for all $n = 1, \ldots, N$, except for the observed nodes $o_{o_1} = o_{o_2} = \cdots = o_{o_{t-1}} = 1$. Finally, define the vector $\vec{d}$, with $d_n = \infty$ for all $n = 1, \ldots, N$, except for $d_i = 0$. This vector represents the distance of a generic node from node $i$. Distance is measured in terms of the value of pairwise entropies along the path.

1. Select $n = \arg\min_{m|v_m<0} d_m$, i.e., the node at minimal distance from $i$ among those not yet visited.

2. Look at all neighbors of node $n$. For a given neighbor $m$ that has not yet been visited, i.e., $v_m < 0$, apply one of the following mutually exclusive operations:

   a) This part applies to an observed node $m$ that we didn't have yet considered, thus we have $o_m > 0$, $s_m < 0$ and $d_m = \infty$. First decide the value of the variable $s_m$. Set $s_m = n$ if $o_n > 0$ or $n == i$, while set $s_m = s_n$, otherwise. Finally compute $d_m = d_n + \mathcal{H}(m|s_m)$.

   b) This part applies to an observed node $m$ that we have already considered, but for which we may find a shorter path towards $i$, thus we have $o_m > 0$, $s_m > 0$ and $d_m < \infty$. We distinguish two subcases. If $o_n > 0$ or $n == i$ and $d_n + \mathcal{H}(m|n) < d_m$, then set $s_m = n$ and recalculate $d_m = d_n + \mathcal{H}(m|s_m)$. Otherwise, if $o_n < 0$ and $d_n + \mathcal{H}(m|s_n) < d_m$, then set $s_m = s_n$ and recalculate $d_m = d_n + \mathcal{H}(m|s_m)$.

   c) This part applies to a nonobserved node $m$, thus we require $o_m < 0$. We set $d_m = d_n$. If $o_n < 0$, set $s_m = s_n$, otherwise set $s_m = n$.

3. Set $v_n = 1$, and go back to point 1 until all nodes are marked as visited.

At the end of the algorithm, we can compute

$$\mathcal{H}_{pair}(o_1, \ldots, o_{t-1}|i) = \sum_{j=1}^{t-1} \mathcal{H}(o_j|s_{o_j}) \,.$$

A Python implementation of the algorithm can be found at `https://github.com/filrad/Maximum-Entropy-Sampling`.

## Numerical simulations of stochastic processes

In all processes we consider, we sample $T$ microscopic configurations $(x_1, x_2, \ldots, x_N)$ of the system by means of Monte Carlo simulations. Simulations are run on a given graph $G$ and for a given set of parameters of the stochastic process. The $T$ sampled configurations are used to compute the marginal entropy $\mathcal{H}(i)$ for every node $i$, and the pairwise conditional entropy $\mathcal{H}(i|j)$ for every pair of nodes $i$ and $j$ in the graph. In our simulations, we set at least $T = 1,000$.

### Ising model

To sample a configuration, we make use of a standard Metropolis-Hastings algorithm [3] with fixed temperature $1/\beta$ and external magnetic field $h = 1/N$, where $N$ is the size of the graph. We let the system evolve for a number of spin flips equal to $1,000\,N$.

### Independent Cascade (IC) model

To simulate model dynamics [4], we start from a configuration where all nodes are in the $S$ state, except for one randomly chosen node in state $I$. We then run Monte Carlo simulations with fixed value of the probability of spreading $p$. We let the system reach a static configuration where no infected nodes are present. We finally set $x_i = 1$ if node $i$ is in state $R$, or $x_i = 0$, otherwise.

### Standard Susceptible-Infected-Susceptible (SIS) model

We start from an initial configuration where all nodes are in the $I$ state. We then run a Gillespie algorithm [5] with $\mu = 1$ and fixed value of $\lambda$. We consider a maximum number of iterations equal to $100\,N$, with $N$ size of the network. In this way we consider active metastable configurations above the epidemic transition. We then set for every node $i$ in the graph $x_i = 1$ if node $i$ is state $I$, or $x_i = 0$, otherwise.

### Modified Susceptible-Infected-Susceptible (MSIS) model

Simulations of the model originally considered in Ref. [6] are performed exactly as described for the standard SIS.

## Analysis of trees

We illustrate results of a systematic analysis of the different stochastic processes running on a tree $G$ with $N = 100$. The tree is an instance of the De Solla-Price model [7], with fitness parameter $a = 1.2$ and number of new connections per node $c = 1$, obtained by neglecting directions on the edges. The same instance of the model is used in all our results reported here. Qualitatively similar results can be obtained for any other graph that is a tree.

First, we perform a test of dependence among variables in the tree $G$. We compute

$$\langle \Delta \mathcal{H}_p \rangle = \frac{\sum_{i,j} A_{i,j} \left[ \mathcal{H}(i) - \mathcal{H}(i|j) \right]}{\sum_{i,j} A_{i,j}} \; , \tag{SM5}$$

that is the average value of the difference between marginal and pair-wise entropy among pairs of physically connected nodes. $A$ is the adjacency matrix of $G$, with generic element $A_{i,j} = A_{j,i} = 1$ if node $i$ and $j$ are connected, while $A_{i,j} = A_{j,i} = 0$, otherwise. The sums in the numerator and denominator of the expression of Eq. (SM5) run over all ordered pairs of distinct nodes. We consider the behavior of the above quantity as a function of the number $T$ of sampled configurations. As $\mathcal{H}(i) = \mathcal{H}(i|j)$ only if $i$ is independent of $j$, we expect to see $\langle \Delta \mathcal{H}_p \rangle \to_{T \to \infty} 0$ for a stochastic process with independent variables. We should have instead $\langle \Delta \mathcal{H}_p \rangle \to_{T \to \infty} c > 0$ in presence of dependence among variables. Second, we compute the quantity

$$\langle \Delta \mathcal{H}_t \rangle = \frac{\sum_{i,j,k} A_{i,k} A_{j,k} \left[ \mathcal{H}(i|k) + \mathcal{H}(j|k) - \mathcal{H}(i,j|k) \right]}{\sum_{i,j,k} A_{i,k} A_{j,k}} \; . \tag{SM6}$$

This quantity allows us to establish whether variables satisfy conditional independence with respect to $G$. Please note that $G$ is a tree, so that $\langle \Delta \mathcal{H}_t \rangle$ represents the average value, over all the triplets of nodes $i$, $j$ and $k$ connected as $i - k - j$, of the difference among disjoint and joint entropy of the nodes $i$ and $j$ conditioned to the node $k$. $\langle \Delta \mathcal{H}_t \rangle \to_{T \to \infty} 0$ indicates conditional independence. An asymptotic value of $\langle \Delta \mathcal{H}_t \rangle$ larger than zero indicates instead that conditional independence is violated.

As fig. SM3b shows for the Ising model, variables associated to neighboring nodes are dependent one on the other. Further, numerical results for the Ising model confirm that the model satisfies conditional independence, as shown in fig. SM3c. We report results obtained for different sampling strategies and entropy approximations in figs. SM3d-f. The amount of independence among neighboring variables is a good indicator of the improvement obtained by using our entropy approximation over the individual-based mean-field approximation, as the difference between $\mathcal{H}_{ind}(\mathbb{O}_{ind})$ and $\mathcal{H}_{pair}(\mathbb{O}_{ind})$ shows. We further have $\mathcal{H}_{pair}(\mathbb{O}_{pair}) - \mathcal{H}_{pair}(\mathbb{O}_{ind}) > 0$, stating that our greedy strategy for the selection of the best set of observed nodes outperforms the mere strategy based on the selection of the nodes with top marginal entropy.

The same conclusions are valid for the MSIS model. Results are reported in fig. SM4. Qualitative considerations are identical to those valid for the Ising model. We note however that this model reaches an equilibrium configuration where variables are almost independent one on the other, as the results of fig. SM4 demonstrate. As a result, using our approximation doesn't represent a significant improvement over the individual-node mean-field approximation.

Finally, we report the results for the two out-of-equilibrium processes: the standard SIS in fig. SM5 and the IC model in fig. SM6. The distributions of microscopic variables of both models violate conditional independence. Nonetheless, we note that considerations valid for equilibrium systems are still valid for results that regard entropy approximations and sampling techniques.

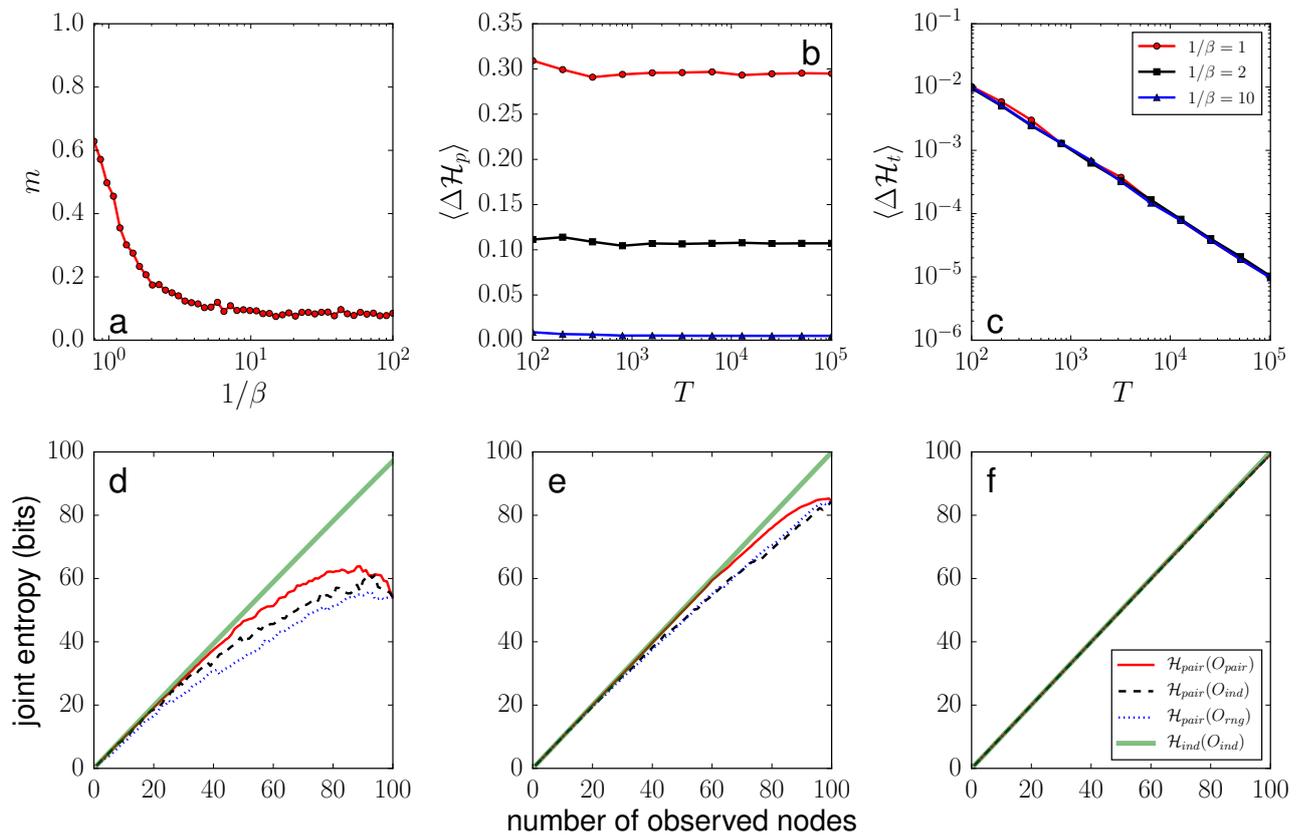

Figure SM3. Results for the Ising model. (a) Magnetization $m$ as a function of the temperature $1/\beta$. (b) Independence test. We plot the quantity defined in Eq. (SM5) as a function of the total number of samples $T$ of equilibrium configurations. Different colors and symbols correspond to different values of $1/\beta$. (c) Conditional independence test. We plot the quantity defined in Eq. (SM6) as a function of the total number of samples $T$ of equilibrium configurations. Different colors and symbols correspond to different values of $1/\beta$. (d-f) Joint entropy of subsets of observed nodes. Different colors and styles of the lines correspond to different choices of the approximation used to estimate entropy, and different subsets of observed nodes. The various panels report results corresponding to different values of $1/\beta$: (d) $1/\beta = 0.5$, (e) $1/\beta = 1.0$, and (f) $1/\beta = 1.5$.

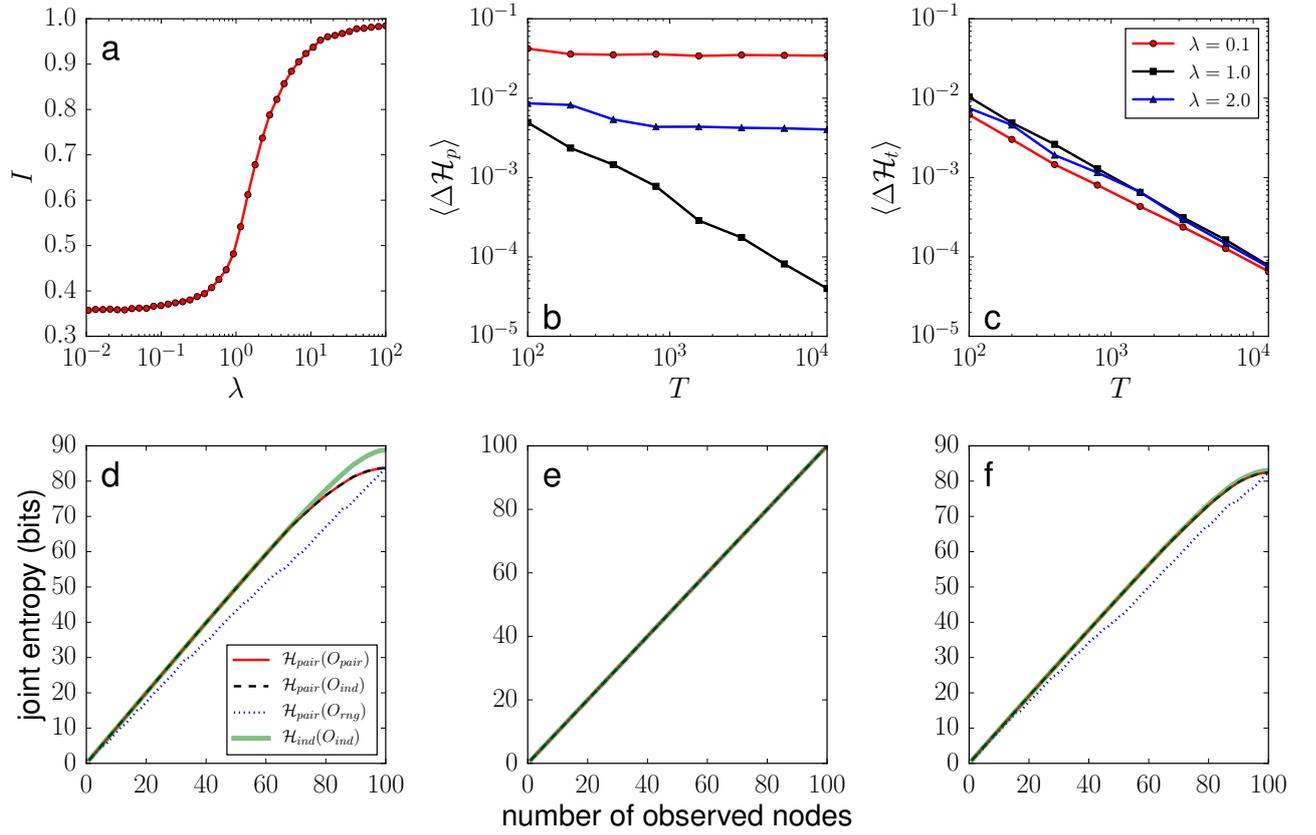

Figure SM4. Results for the MSIS model. (a) Fraction of infected nodes $I$ as a function of the epidemic rate $\lambda$. (b) Independence test. We plot the quantity defined in Eq. (SM5) as a function of the total number of samples $T$ of equilibrium configurations. Different colors and symbols correspond to different values of $\lambda$. (c) Conditional independence test. We plot the quantity defined in Eq. (SM6) as a function of the total number of samples $T$ of equilibrium configurations. Different colors and symbols correspond to different values of $\lambda$. (d-f) Joint entropy of subsets of observed nodes. Different colors and styles of the lines correspond to different choice of the approximation used to estimate entropy, and different subsets of observed nodes. The different panels report results corresponding to different values of $\lambda$: (d) $\lambda = 0.1$, (e) $\lambda = 1.0$, and (f) $\lambda = 2.0$.

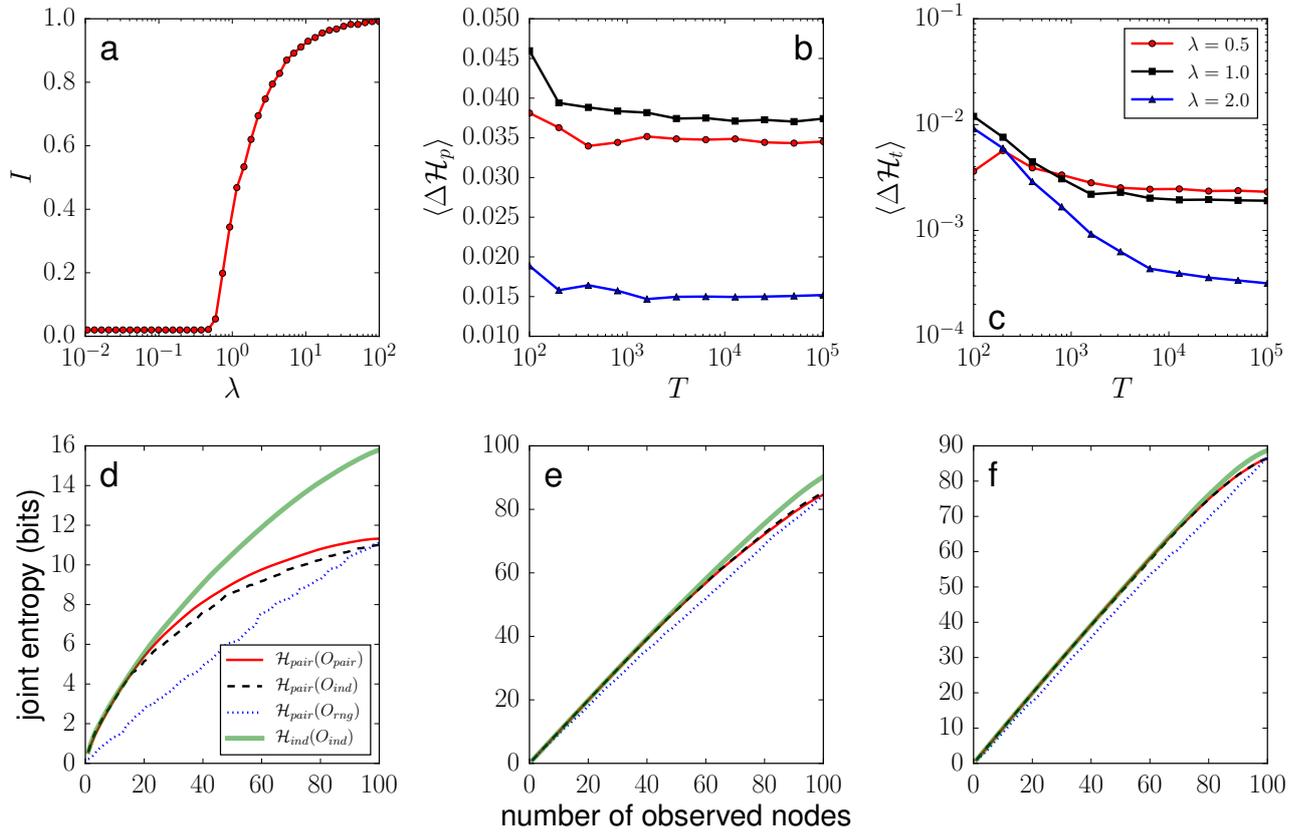

Figure SM5. Results for the standard SIS model. (a) Fraction of infected nodes $I$ as a function of the epidemic rate $\lambda$. (b) Independence test. We plot the quantity defined in Eq. (SM5) as a function of the total number of samples $T$ of quasi-stationary configurations. Different colors and symbols correspond to different values of $\lambda$. (c) Conditional independence test. We plot the quantity defined in Eq. (SM6) as a function of the total number of samples $T$ of quasi-stationary configurations. Different colors and symbols correspond to different values of $\lambda$. (d-f) Joint entropy of subsets of observed nodes. Different colors and styles of the lines correspond to different choice of the approximation used to estimate entropy, and different subsets of observed nodes. The different panels report results corresponding to different values of $\lambda$: (d) $\lambda = 0.1$, (e) $\lambda = 1.0$, and (f) $\lambda = 2.0$.

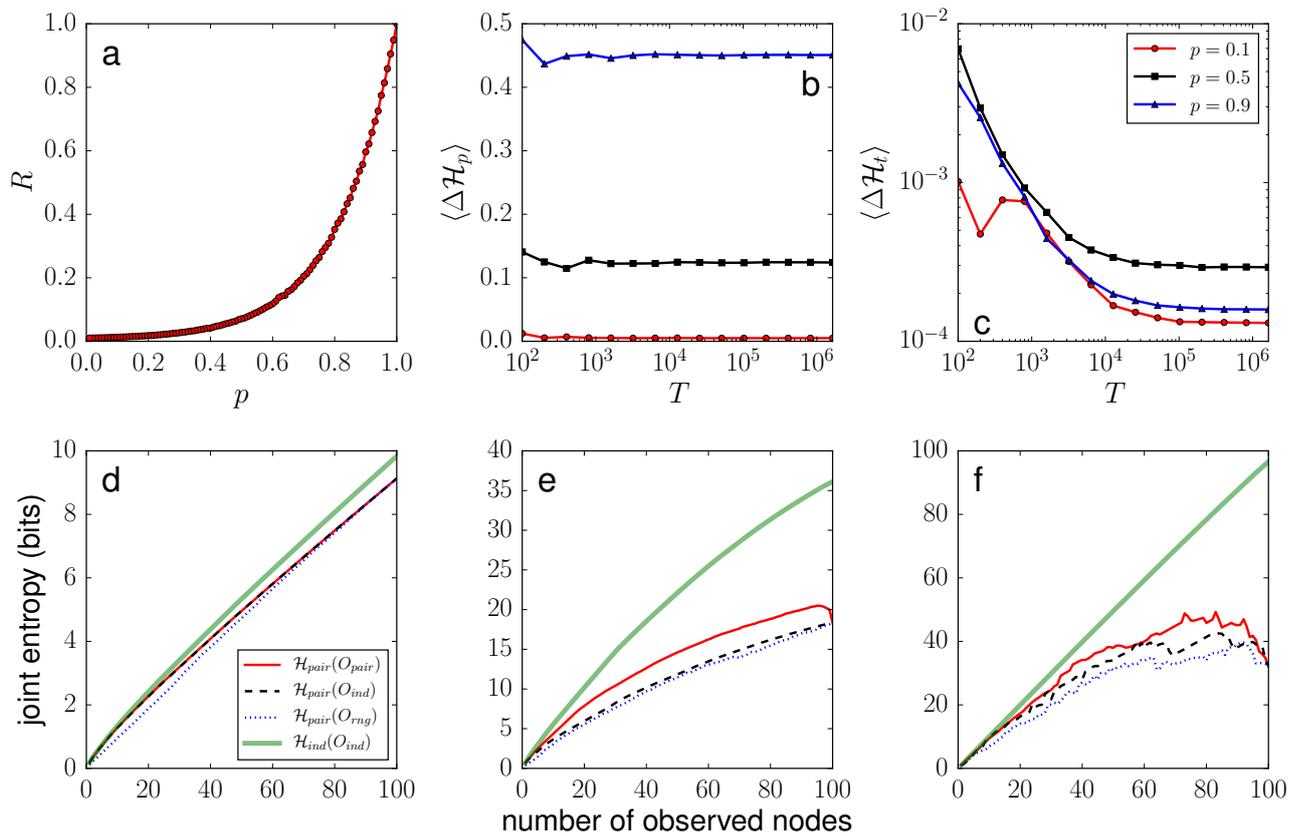

Figure SM6. Results for the IC model. (a) Relative size of the outbreak $R$ as a function of the spreading probability $p$. (b) Independence test. We plot the quantity defined in Eq. (SM5) as a function of the total number of samples $T$ of stationary configurations. Different colors and symbols correspond to different values of $p$. (c) Conditional independence test. We plot the quantity defined in Eq. (SM6) as a function of the total number of samples $T$ of stationary configurations. Different colors and symbols correspond to different values of $p$. (d-f) Joint entropy of subsets of observed nodes. Different colors and styles of the lines correspond to different choice of the approximation used to estimate entropy, and different subsets of observed nodes. The different panels report results corresponding to different values of $p$: (d) $p = 0.1$, (e) $p = 0.5$, and (f) $p = 0.9$.

**Analysis of Networks**

We considered the following networks in our analysis:

- Uncorrelated Configuration Model (UCM). We considered a single instance of the uncorrelated configuration model [8]. The degree distribution of the network is $P(k) \sim k^{-5/2}$ for $k \in [3, \sqrt{N}]$, and $P(k) = 0$, otherwise. The size of the network is $N = 1,000$. The same model has been used to generate the networks considered in Fig. SM2.

- US Air Transportation Network (USATN). This network has size $N = 500$ and was originally considered in Ref. [9].

- Email Communication Network (ECN). The network has size $N = 1,133$ and was originally considered in Ref. [10].

- C. Elegans Neural Network (CENN). The network has size $N = 297$ and was originally considered in Ref. [11].

- Political Blogs Network (PBN). The network has size $N = 1,222$ and was originally considered in Ref. [12].

- Autonomous System Network (ASN). The network has size $N = 6,474$ and was originally considered in Ref. [13].

## Results for the Ising model

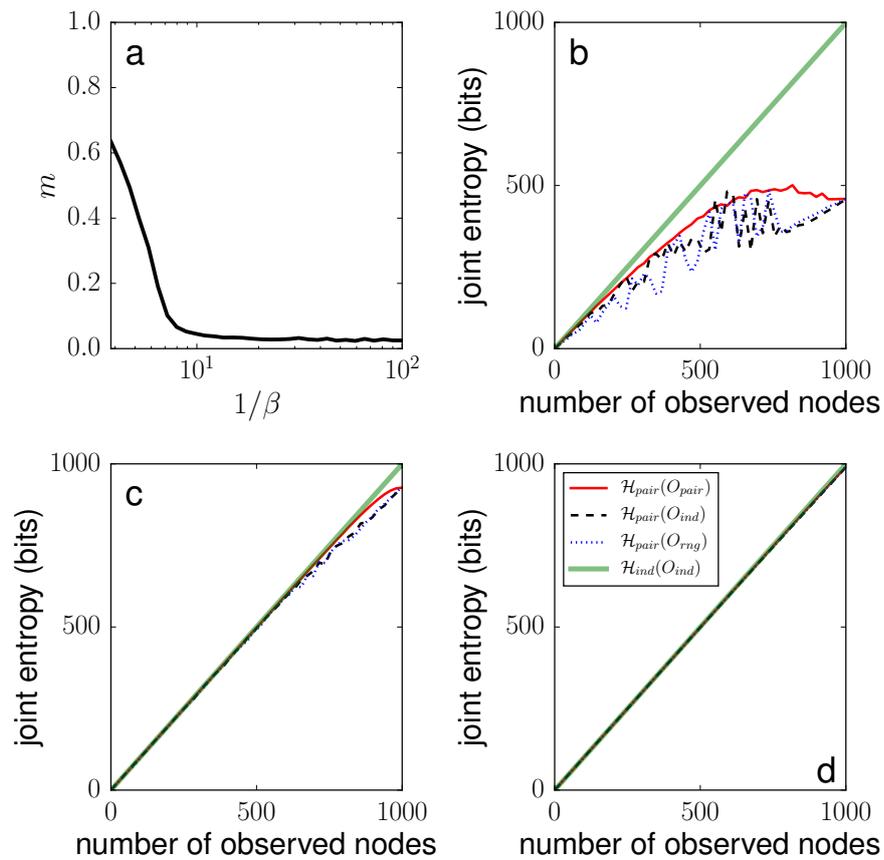

Figure SM7. We consider the Ising model applied to the UCM. The description of the figure is identical to the one of Fig. 1 of the main text. The values of the temperature $1/\beta$ for panels b, c, and d are: b) $1/\beta = 3$, c) $1/\beta = 6$, d) $1/\beta = 10$.

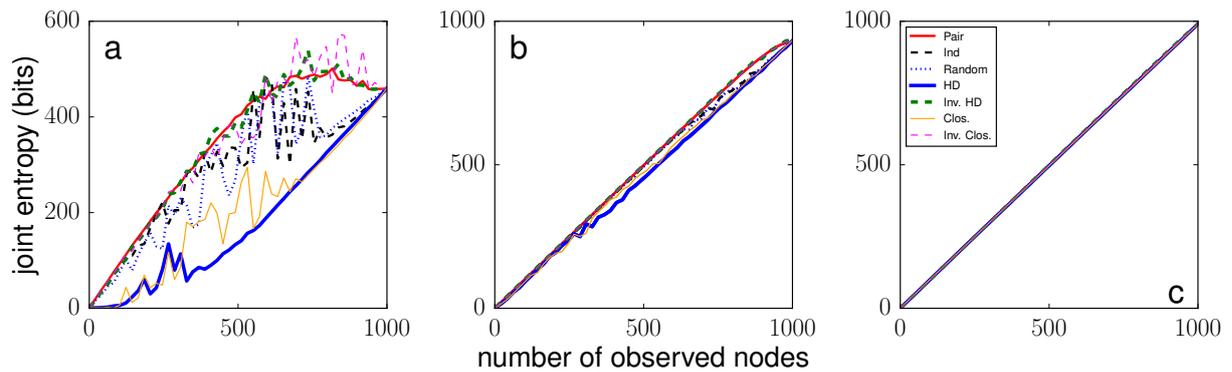

Figure SM8. Ising model applied to the UCM. Panels a, b, and c correspond to the same parameters values as in panels b, c, and d of Fig. SM7, respectively. In all cases, joint entropy is measured with our approximation $\mathcal{H}_{pair}$. We consider different strategies $X$ to construct the observed set $\mathbb{O}_X$. In addition to the same already considered in Fig. SM7, we consider here also sampling strategies based on degree ($X = $ HD), and closeness centrality ($X = $ Clos.), where nodes are ranked according to those metrics. We consider also the inverse strategies, where nodes are ranked in increasing order based on the value of these metrics ($X = $ Inv. HD, and $X = $ Inv. Clos., respectively).

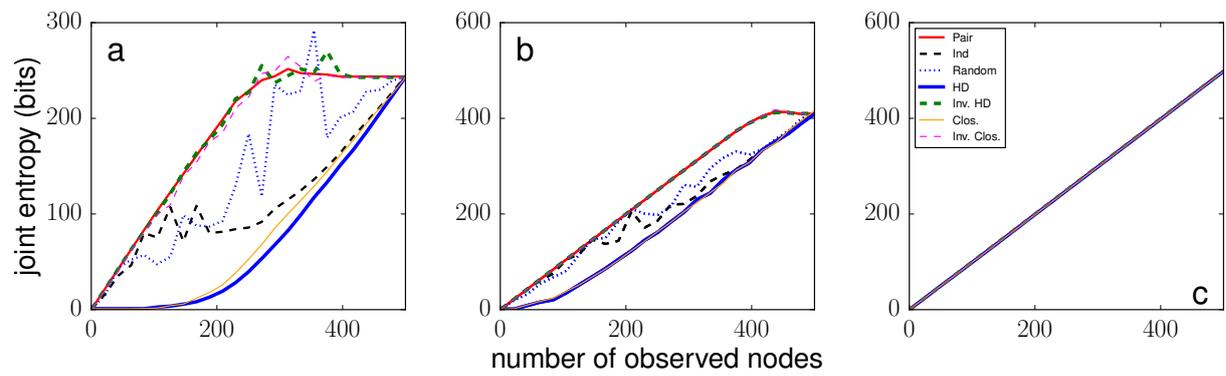

Figure SM9. Ising model applied to the USATN. Panels a, b, and c correspond to the same parameters values as in panels b, c, and d of Fig. 1 of the main text, respectively. The description of the figure is ideintical to the one of Fig. SM8.

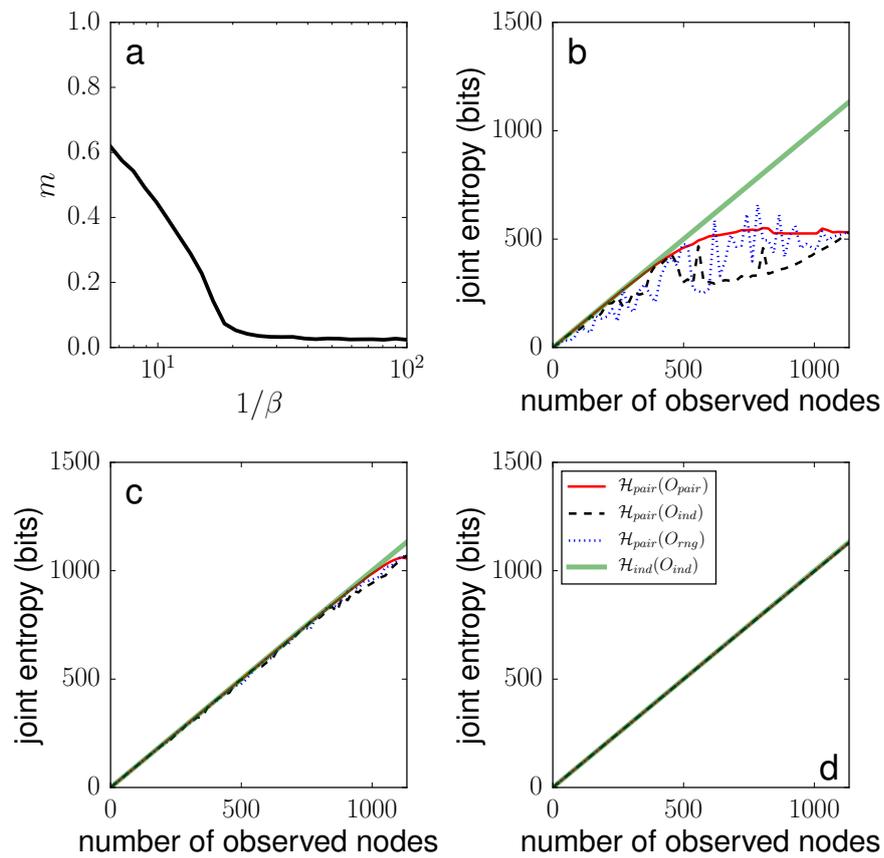

Figure SM10. We consider the Ising model applied to the ECN. The description of panels a–d is the same as of Fig. SM7. The values of the temperature $1/\beta$ for panels b, c, and d are: b) $1/\beta = 5$, c) $1/\beta = 15$, d) $1/\beta = 30$.

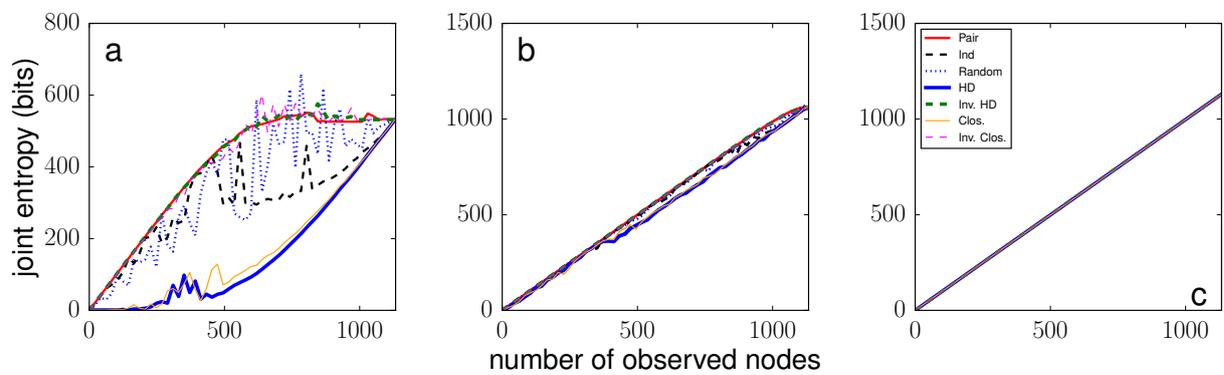

Figure SM11. Ising model applied to the ECN. Panels a, b, and c correspond to the same parameters values as in panels b, c, and d of Fig. SM10, respectively. The description of the figure is ideintical to the one of Fig. SM8.

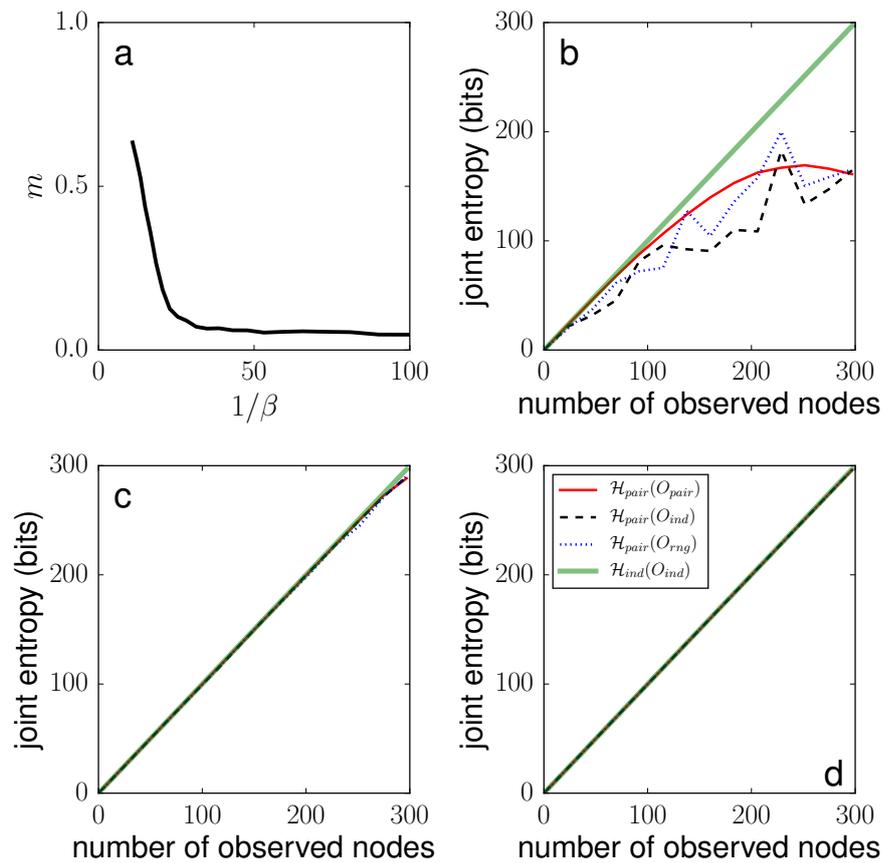

Figure SM12. We consider the Ising model applied to the CENN. The description of panels a–d is the same as of Fig. SM7. The values of the temperature $1/\beta$ for panels b, c, and d are: b) $1/\beta = 10$, c) $1/\beta = 20$, d) $1/\beta = 40$.

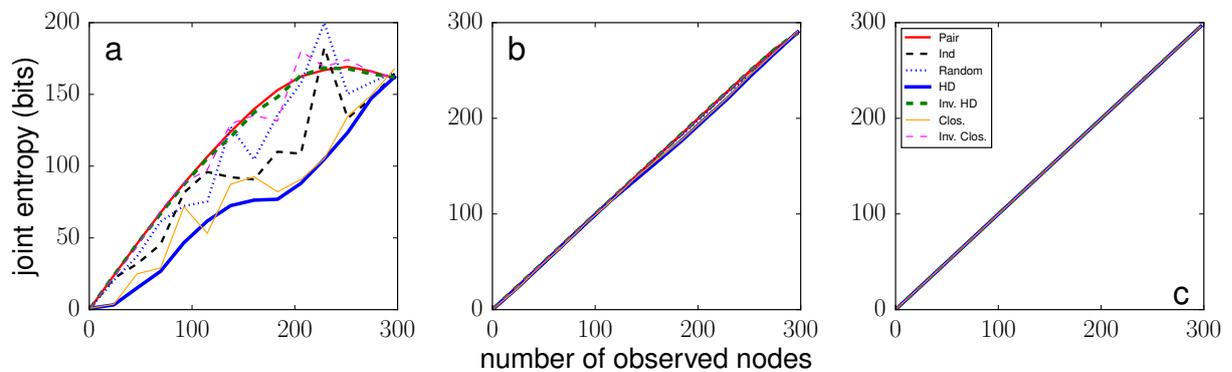

Figure SM13. Ising model applied to the CENN. Panels a, b, and c correspond to the same parameters values as in panels b, c, and d of Fig. SM12, respectively. The description of the figure is ideintical to the one of Fig. SM8.

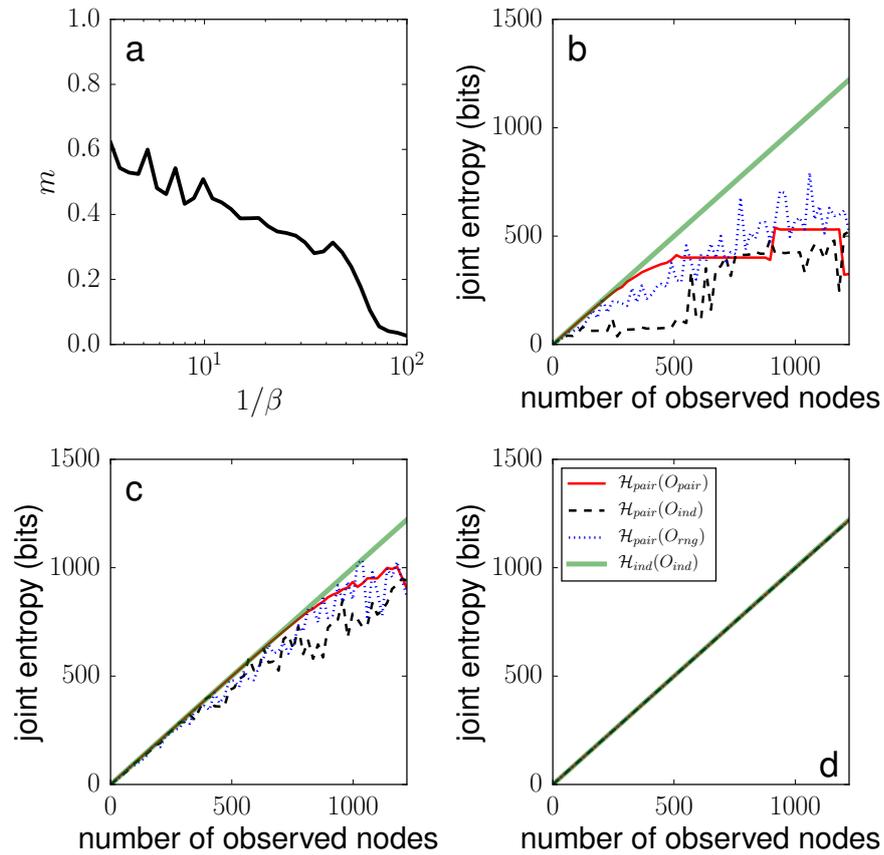

Figure SM14. We consider the Ising model applied to the PBN. The description of panels a–d is the same as of Fig. SM7. The values of the temperature $1/\beta$ for panels b, c, and d are: b) $1/\beta = 3$, c) $1/\beta = 25$, d) $1/\beta = 80$.

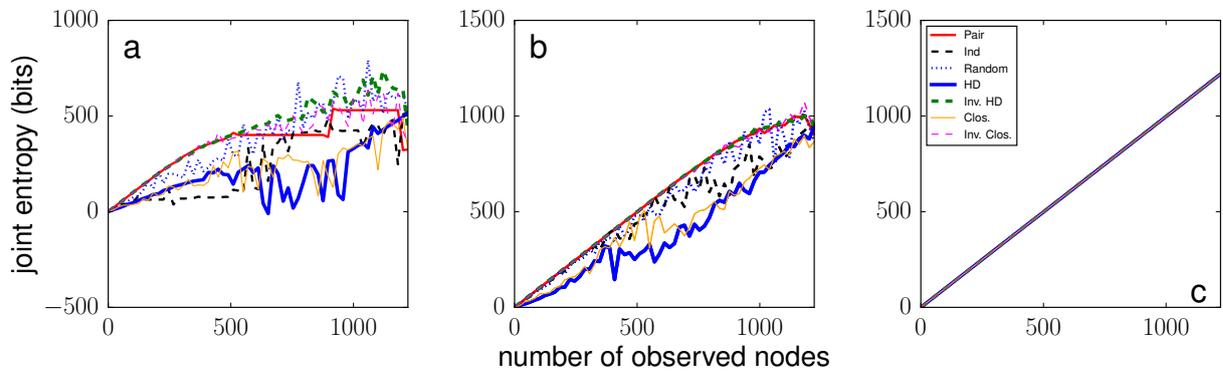

Figure SM15. Ising model applied to the PBN. Panels a, b, and c correspond to the same parameters values as in panels b, c, and d of Fig. SM14, respectively. The description of the figure is ideintical to the one of Fig. SM8.

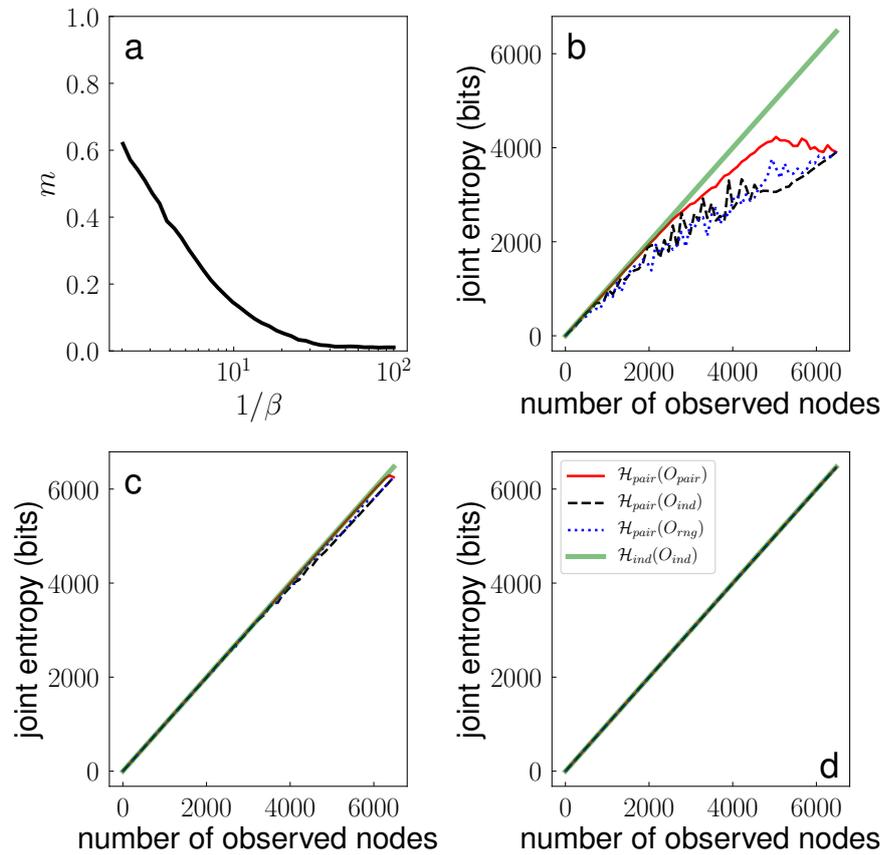

Figure SM16. We consider the Ising model applied to the ASN. The description of panels a–d is the same as of Fig. SM7. The values of the temperature $1/\beta$ for panels b, c, and d are: b) $1/\beta = 2$, c) $1/\beta = 10$, d) $1/\beta = 50$.

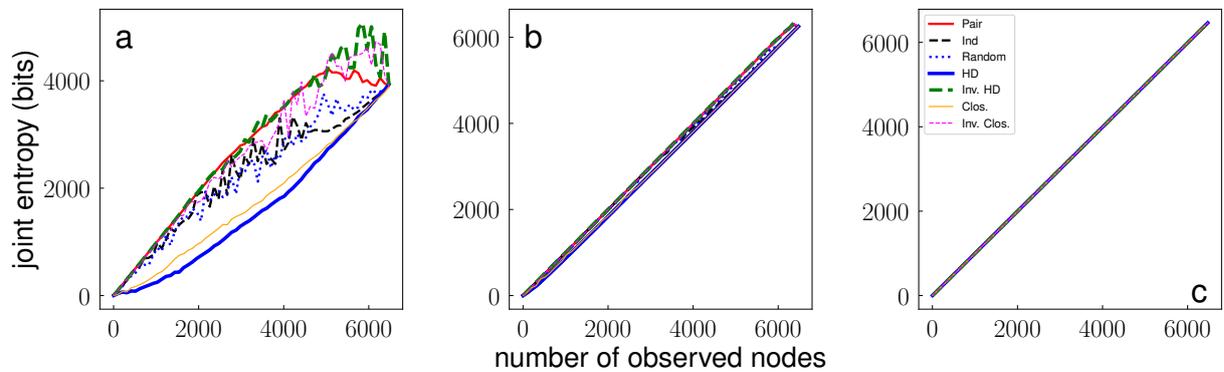

Figure SM17. Ising model applied to the ASN. Panels a, b, and c correspond to the same parameters values as in panels b, c, and d of Fig. SM16, respectively. The description of the figure is ideintical to the one of Fig. SM8.

**Results for the Independent Cascade (IC) model**

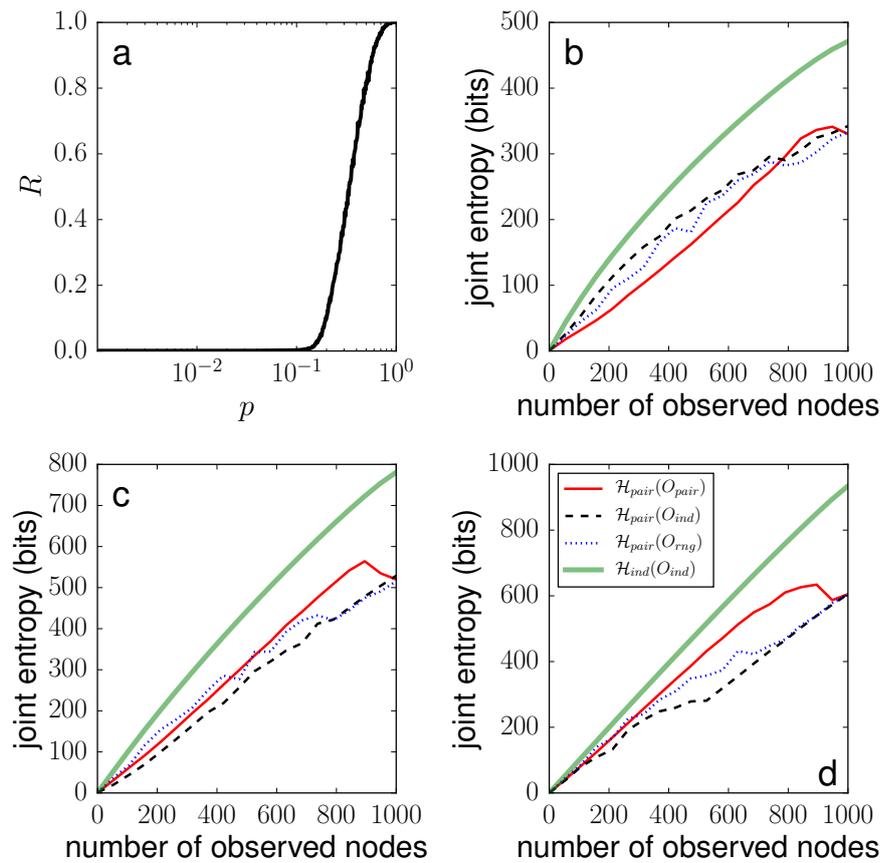

Figure SM18. We consider the IC model applied to the UCM. The description of the figure is identical to the one of Fig. 2 of the main text. The values of the infection probability $p$ for panels b, c, and d are: b) $p = 0.2$, c) $p = 0.25$, d) $p = 0.3$.

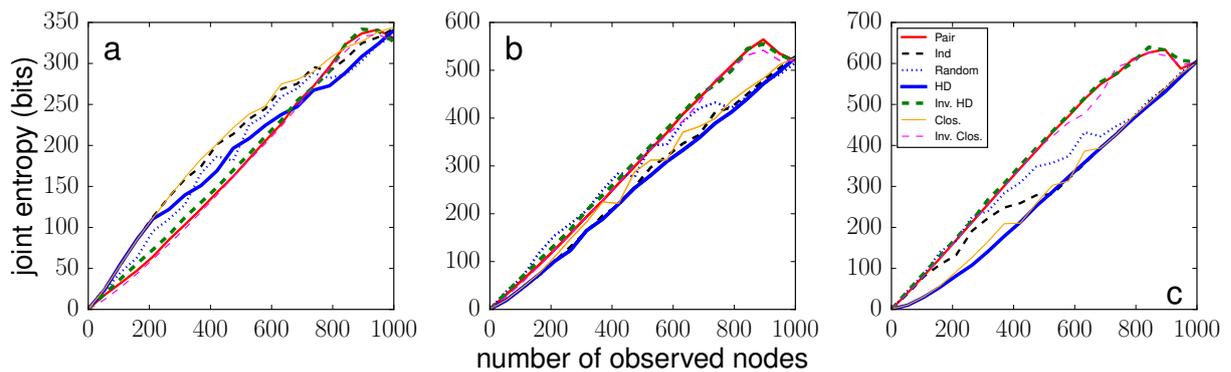

Figure SM19. IC model applied to the UCM. Panels a, b, and c correspond to the same parameters values as in panels b, c, and d of Fig. SM18, respectively. In all cases, joint entropy is measured with our approximation $\mathcal{H}_{pair}$. We consider different strategies $X$ to construct the observed set $\mathbb{O}_X$. In addition to the same already considered in Fig. SM18, we consider here also sampling strategies based on degree ($X = $ HD), and closeness centrality ($X = $ Clos.), where nodes are ranked according to those metrics. We consider also the inverse strategies, where nodes are ranked in increasing order based on the value of these metrics ($X = $ Inv. HD, and $X = $ Inv. Clos., respectively).

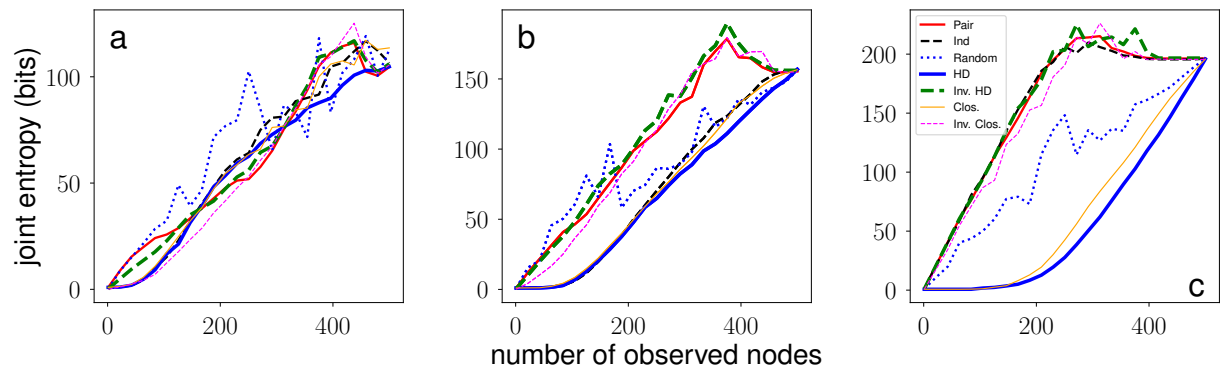

Figure SM20. IC model applied to the USATN. Panels a, b, and c correspond to the same parameters values as in panels b, c, and d of Fig. 2 of the main text, respectively. The description of the figure is ideintical to the one of Fig. SM19.

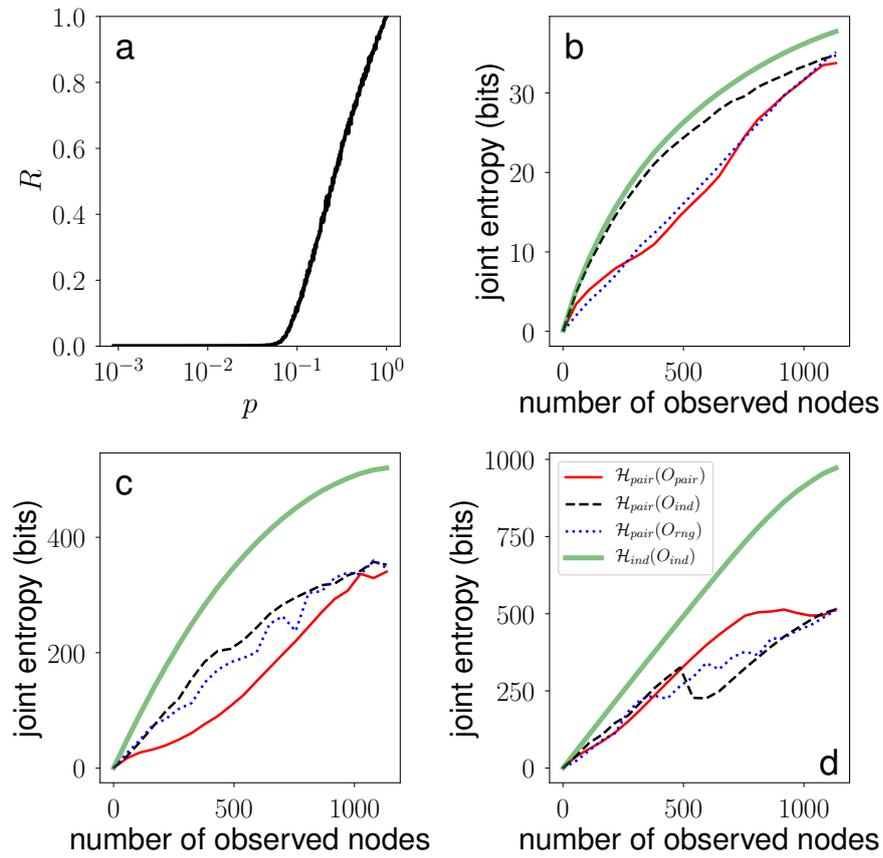

Figure SM21. We consider the IC model applied to the ECN. The description of panels a–d is the same as of Fig. SM18. The values of the infection probability $p$ for panels b, c, and d are: b) $p = 0.05$, c) $p = 0.1$, d) $p = 0.2$.

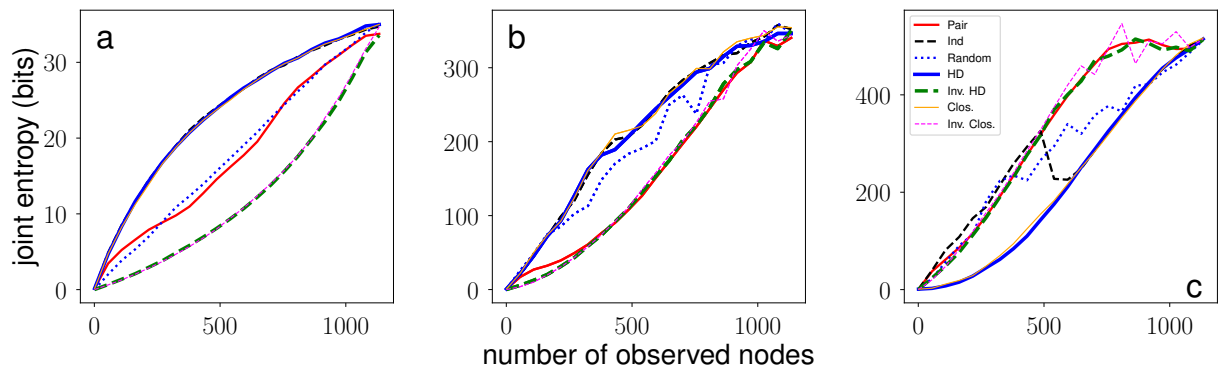

Figure SM22. IC model applied to the ECN. Panels a, b, and c correspond to the same parameters values as in panels b, c, and d of Fig. SM21, respectively. The description of the figure is ideintical to the one of Fig. SM19.

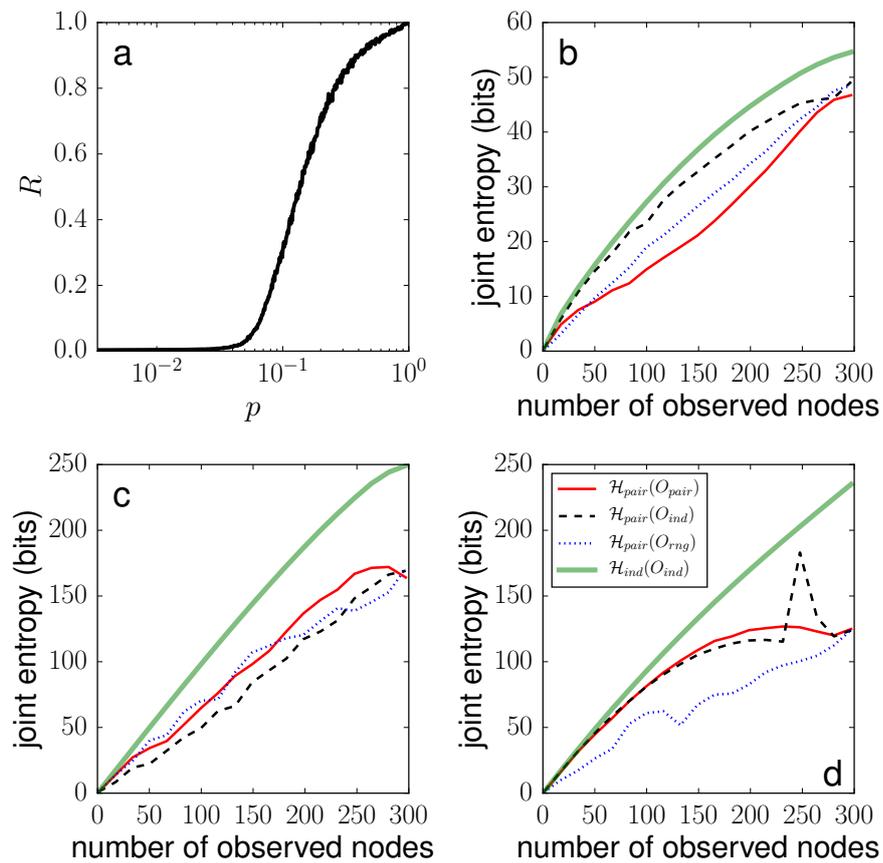

Figure SM23. We consider the IC model applied to the CENN. The description of panels a–d is the same as of Fig. SM18. The values of the infection probability $p$ for panels b, c, and d are: b) $p = 0.05$, c) $p = 0.1$, d) $p = 0.2$.

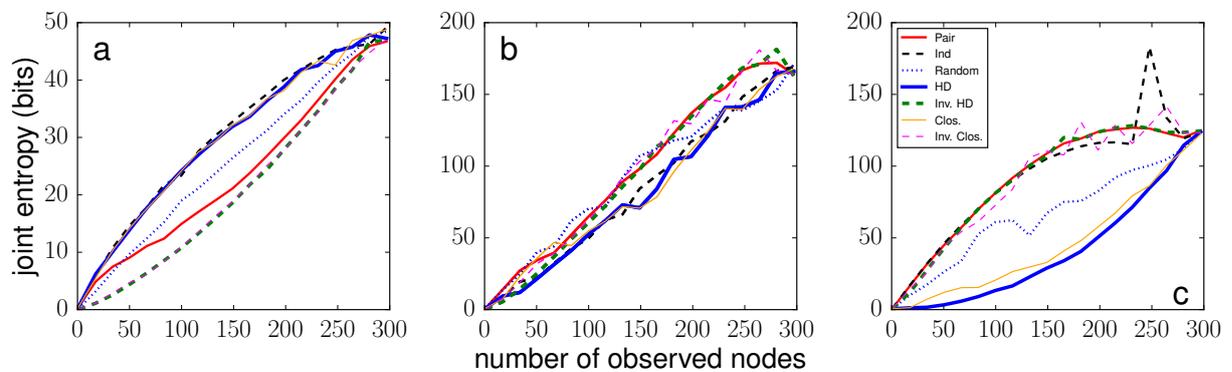

Figure SM24. IC model applied to the CENN. Panels a, b, and c correspond to the same parameters values as in panels b, c, and d of Fig. SM23, respectively. The description of the figure is ideintical to the one of Fig. SM19.

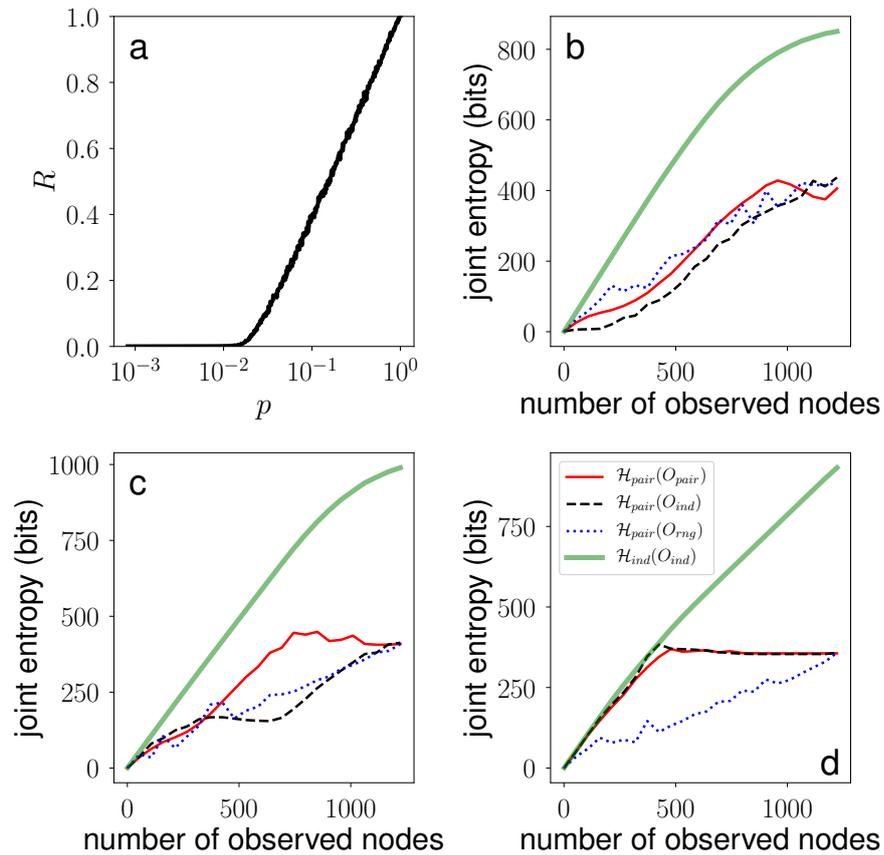

Figure SM25. We consider the IC model applied to the PBN. The description of panels a–d is the same as of Fig. SM18. The values of the infection probability $p$ for panels b, c, and d are: b) $p = 0.06$, c) $p = 0.1$, d) $p = 0.3$.

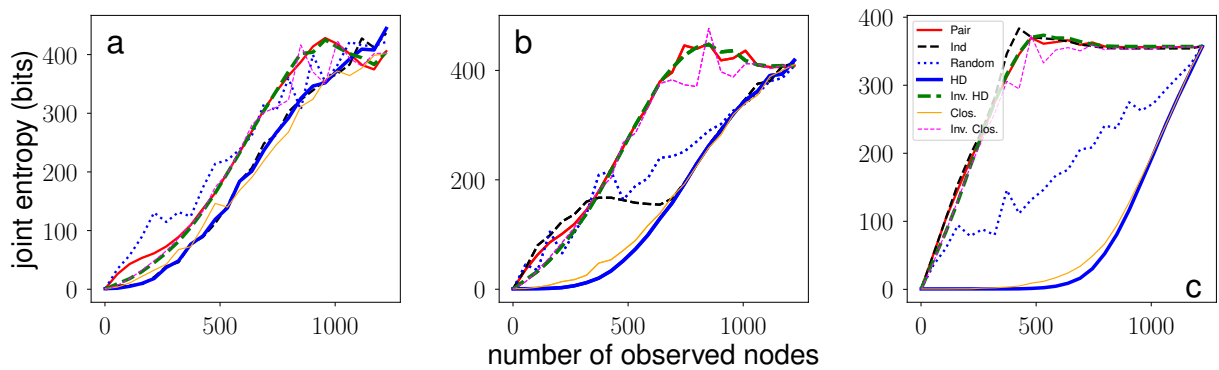

Figure SM26. IC model applied to the PBN. Panels a, b, and c correspond to the same parameters values as in panels b, c, and d of Fig. SM25, respectively. The description of the figure is ideintical to the one of Fig. SM19.

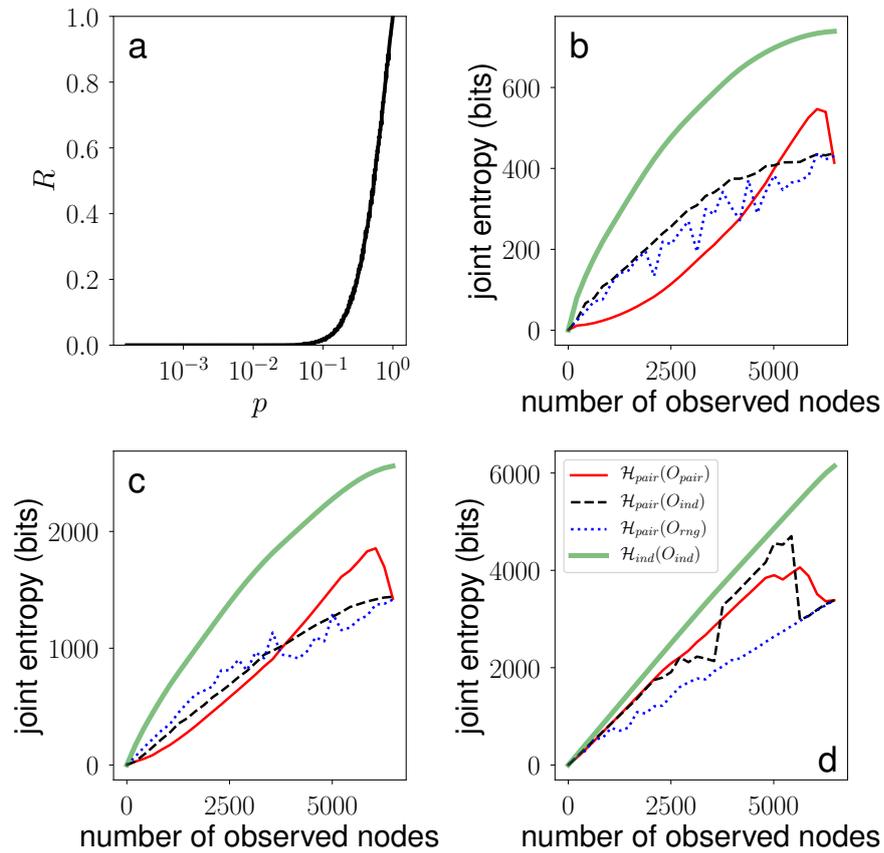

Figure SM27. We consider the IC model applied to the ASN. The description of panels a–d is the same as of Fig. SM18. The values of the infection probability $p$ for panels b, c, and d are: b) $p = 0.1$, c) $p = 0.2$, d) $p = 0.5$.

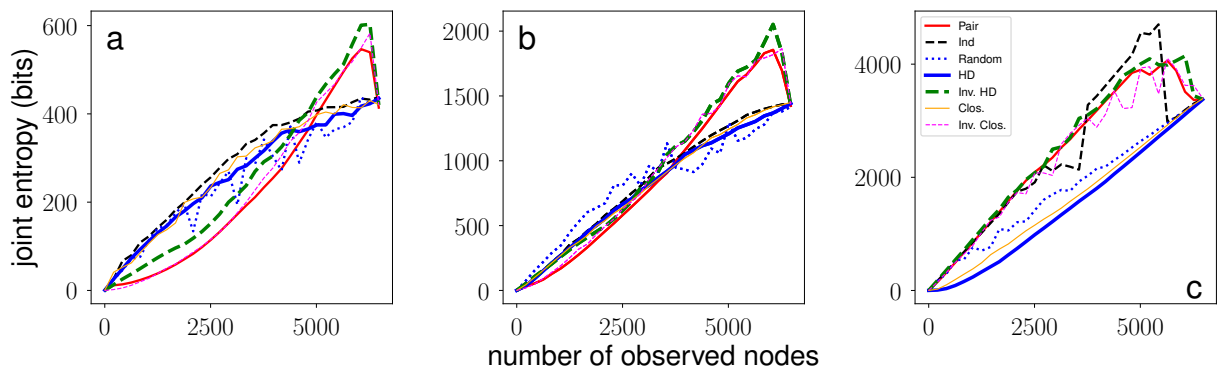

Figure SM28. IC model applied to the ASN. Panels a, b, and c correspond to the same parameters values as in panels b, c, and d of Fig. SM27, respectively. The description of the figure is ideintical to the one of Fig. SM19.

**Results for the standard Susceptible-Infected-Susceptible (SIS) model**

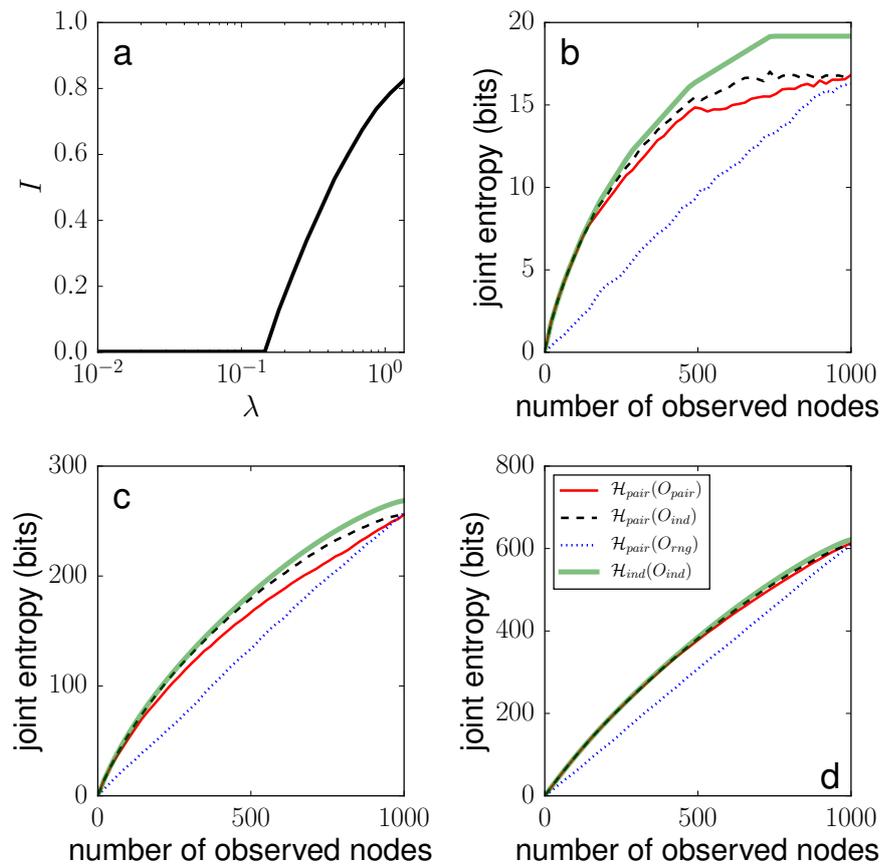

Figure SM29. We consider the SIS model applied to the UCM. a) Fraction of infected nodes $I$ as a function of the infection rate $\lambda$. The description of panels b, c, and d is identical to the one of Fig. 1 of the main text, with the only difference that here we are considering the SIS model for different values of the infection rate $\lambda$. The values of the infection rate $\lambda$ for panels b, c, and d are: b) $\lambda = 0.1$, c) $\lambda = 0.16$, d) $\lambda = 0.2$.

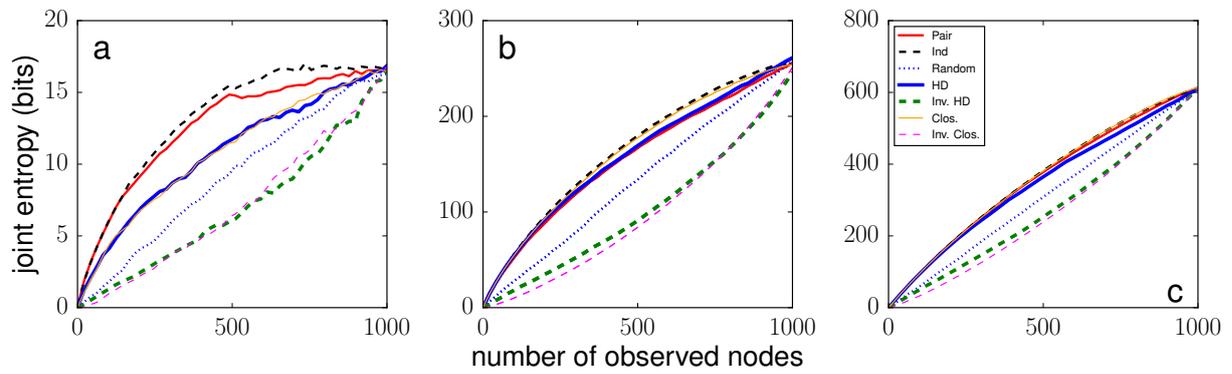

Figure SM30. SIS model applied to the UCM. Panels a, b, and c correspond to the same parameters values as in panels b, c, and d of Fig. SM29, respectively. In all cases, joint entropy is measured with our approximation $\mathcal{H}_{pair}$. We consider different strategies $X$ to construct the observed set $\mathbb{O}_X$. In addition to the same already considered in Fig. SM29, we consider here also sampling strategies based on degree ($X$ = HD), and closeness centrality ($X$ = Clos.), where nodes are ranked according to those metrics. We consider also the inverse strategies, where nodes are ranked in increasing order based on the value of these metrics ($X$ = Inv. HD, and $X$ = Inv. Clos., respectively).

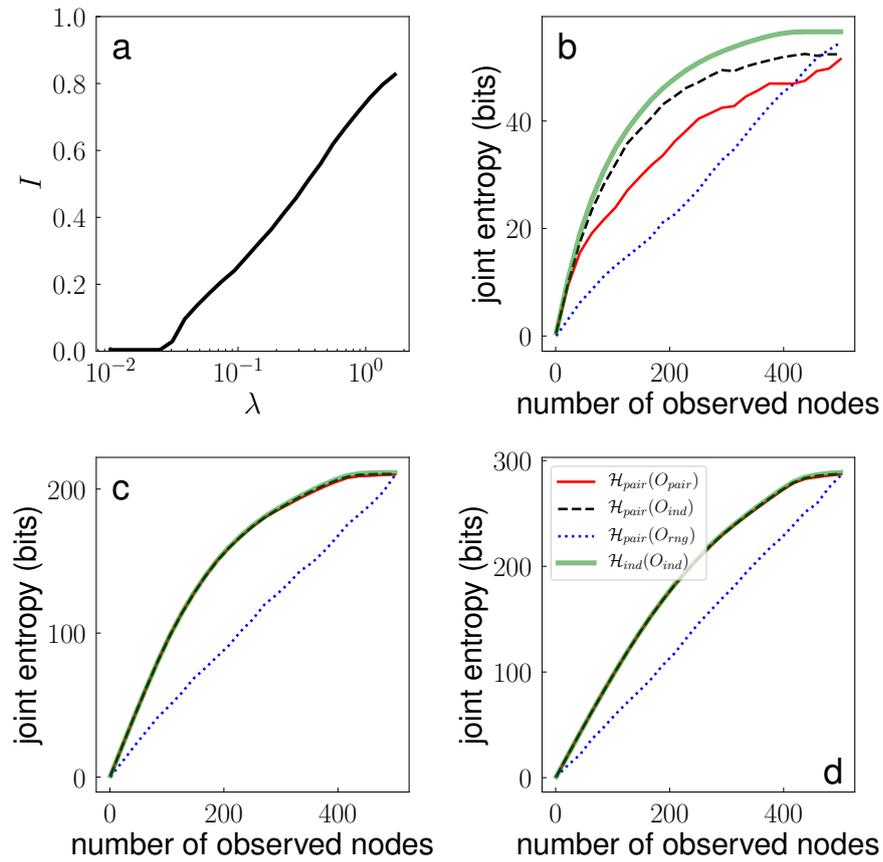

Figure SM31. We consider the SIS model applied to the USATN. The description of panels a–d is the same as of Fig. SM29. The values of the infection rate $\lambda$ for panels b, c, and d are: b) $\lambda = 0.03$, c) $\lambda = 0.05$, d) $\lambda = 0.1$.

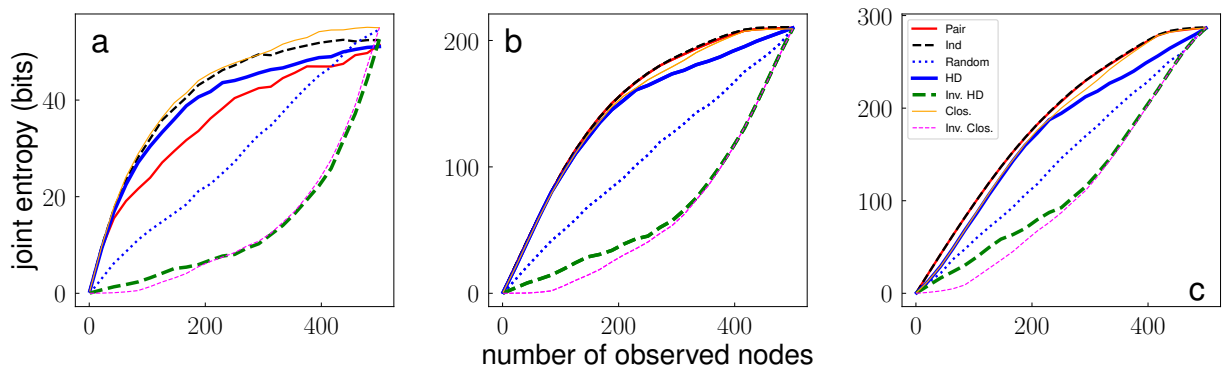

Figure SM32. SIS model applied to the USATN. Panels a, b, and c correspond to the same parameters values as in panels b, c, and d of Fig. SM31, respectively. The description of the figure is ideintical to the one of Fig. SM30.

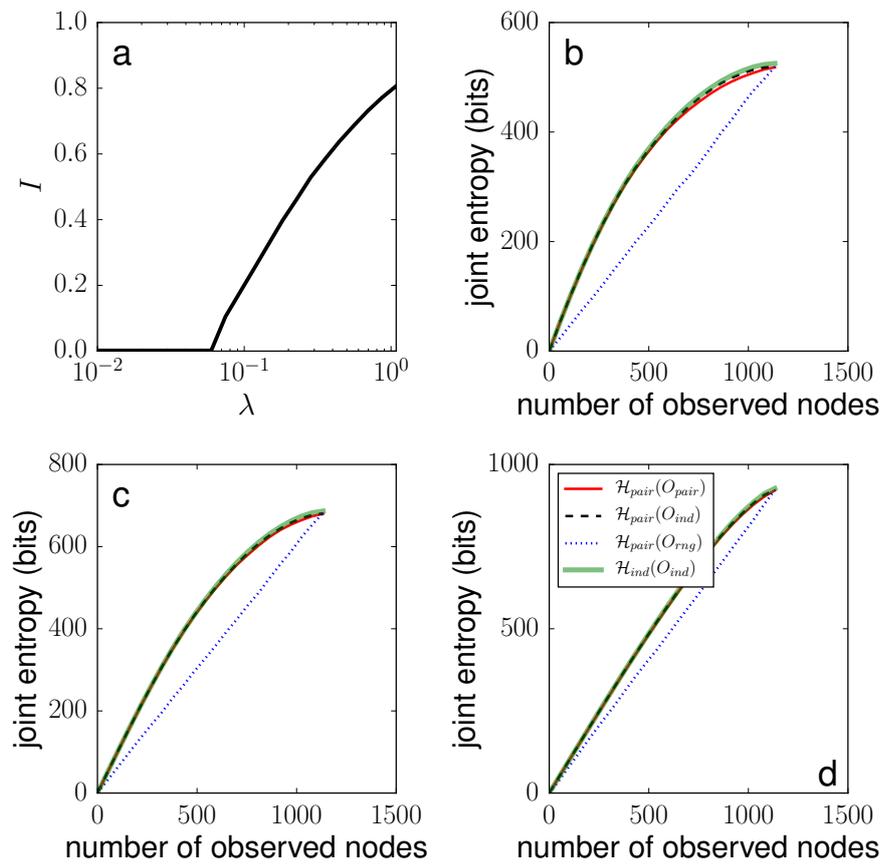

Figure SM33. We consider the SIS model applied to the ECN. The description of panels a–d is the same as of Fig. SM29. The values of the infection rate $\lambda$ for panels b, c, and d are: b) $\lambda = 0.08$, c) $\lambda = 0.1$, d) $\lambda = 0.2$.

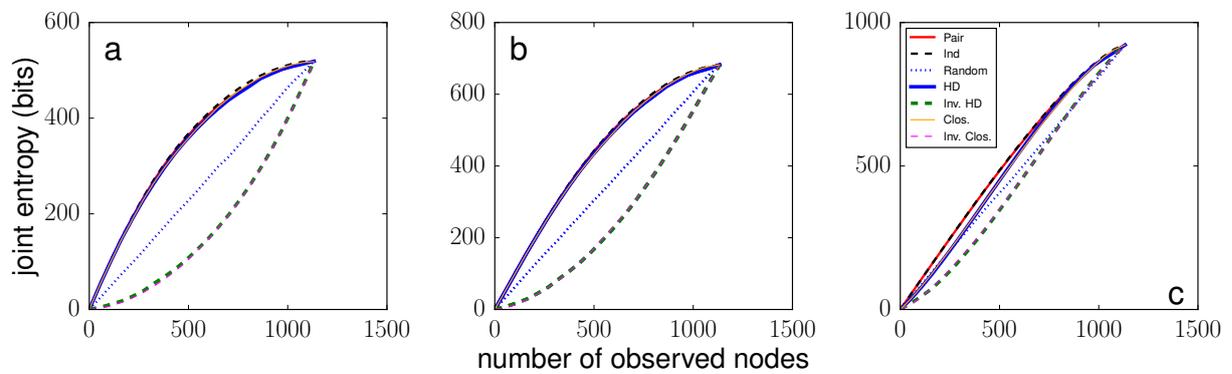

Figure SM34. SIS model applied to the ECN. Panels a, b, and c correspond to the same parameters values as in panels b, c, and d of Fig. SM33, respectively. The description of the figure is ideintical to the one of Fig. SM30.

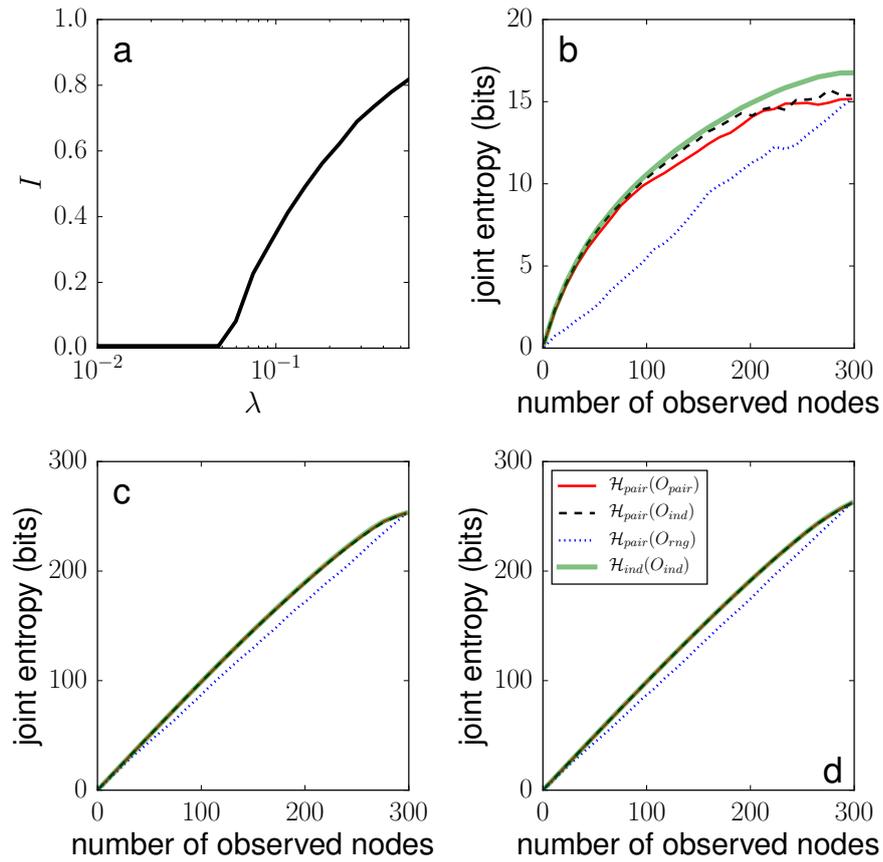

Figure SM35. We consider the SIS model applied to the CENN. The description of panels a–d is the same as of Fig. SM29. The values of the infection rate $\lambda$ for panels b, c, and d are: b) $\lambda = 0.05$, c) $\lambda = 0.1$, d) $\lambda = 0.2$.

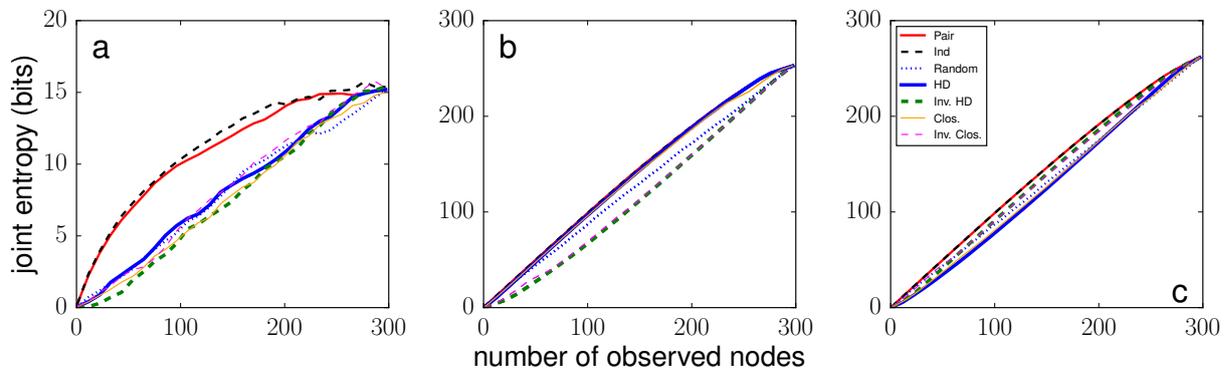

Figure SM36. SIS model applied to the CENN. Panels a, b, and c correspond to the same parameters values as in panels b, c, and d of Fig. SM35, respectively. The description of the figure is ideintical to the one of Fig. SM30.

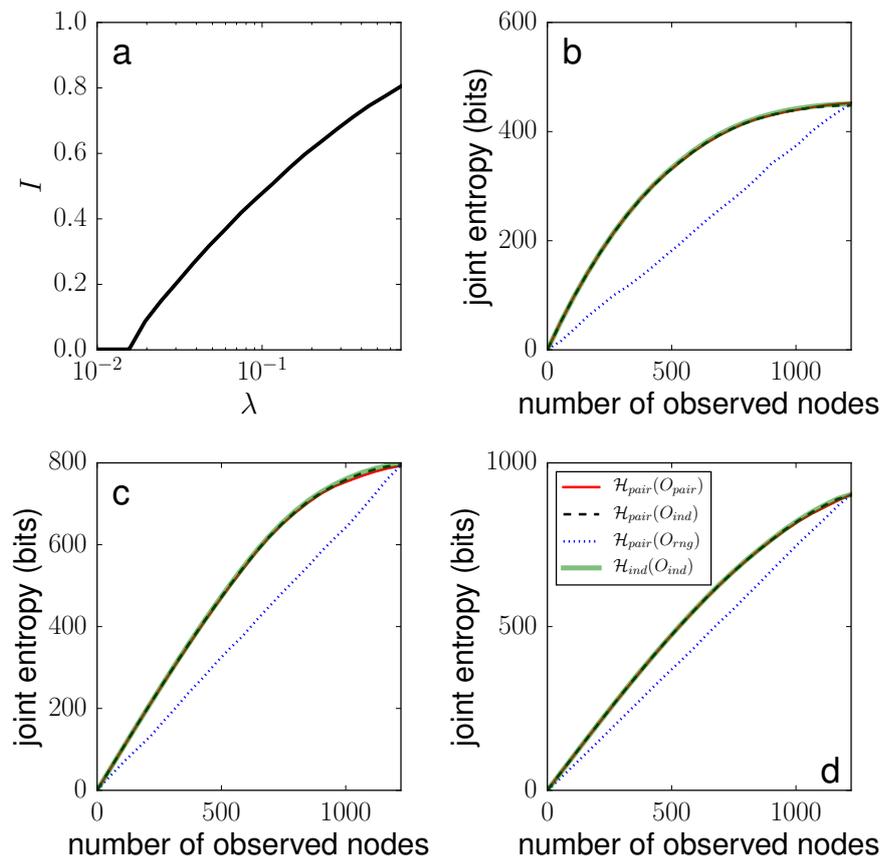

Figure SM37. We consider the SIS model applied to the PBN. The description of panels a–d is the same as of Fig. SM29. The values of the infection rate $\lambda$ for panels b, c, and d are: b) $\lambda = 0.02$, c) $\lambda = 0.04$, d) $\lambda = 0.1$.

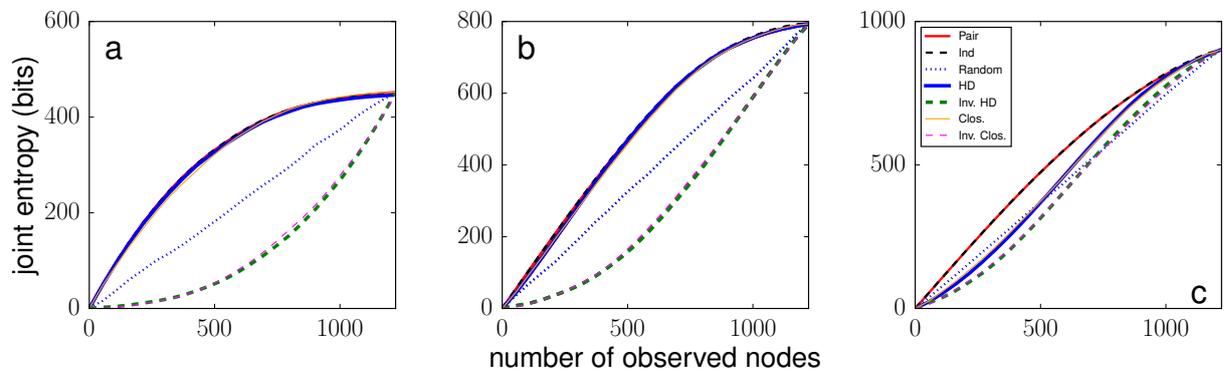

Figure SM38. SIS model applied to the PBN. Panels a, b, and c correspond to the same parameters values as in panels b, c, and d of Fig. SM37, respectively. The description of the figure is ideintical to the one of Fig. SM30.

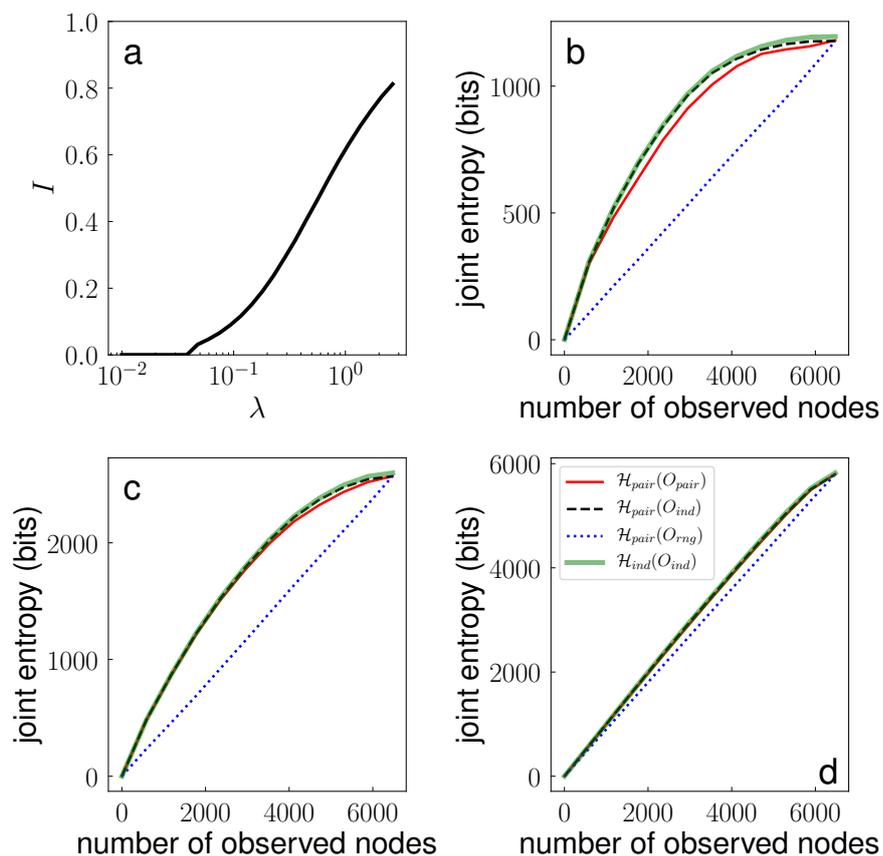

Figure SM39. We consider the SIS model applied to the ASN. The description of panels a–d is the same as of Fig. SM29. The values of the infection rate $\lambda$ for panels b, c, and d are: b) $\lambda = 0.05$, c) $\lambda = 0.1$, d) $\lambda = 1.0$.

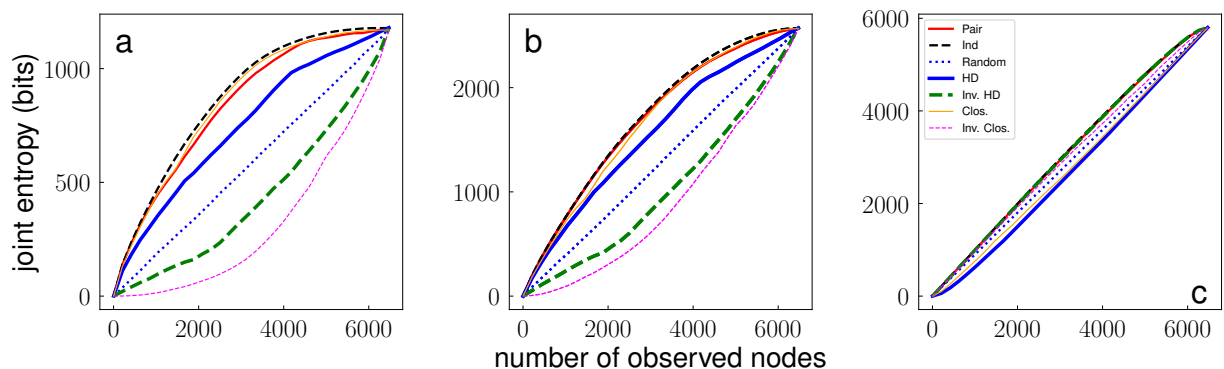

Figure SM40. SIS model applied to the ASN. Panels a, b, and c correspond to the same parameters values as in panels b, c, and d of Fig. SM39, respectively. The description of the figure is ideintical to the one of Fig. SM30.

**Results for the Modified Susceptible-Infected-Susceptible (MSIS) model**

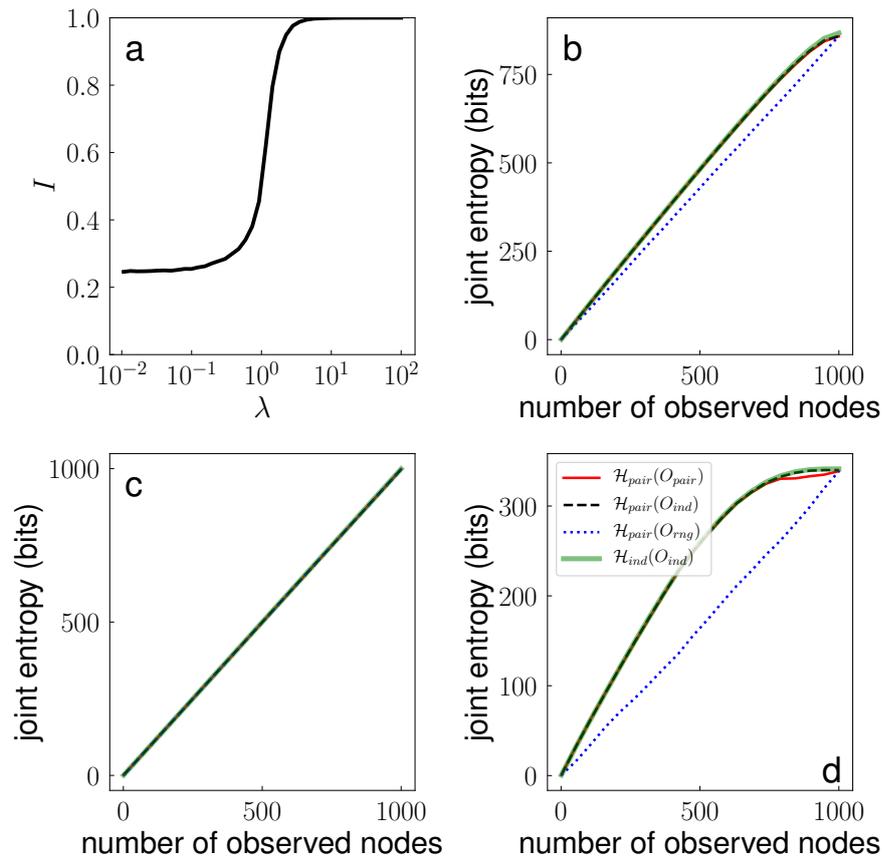

Figure SM41. We consider the MSIS model applied to the UCM. a) Fraction of infected nodes $I$ as a function of the infection rate $\lambda$. The description of panels b, c, and d is identical to the one of Fig. 1 of the main text, with the only difference that here we are considering the MSIS model for different values of the infection rate $\lambda$. The values of the infection rate $\lambda$ for panels b, c, and d are: b) $\lambda = 0.5$, c) $\lambda = 1.0$, d) $\lambda = 2.0$.

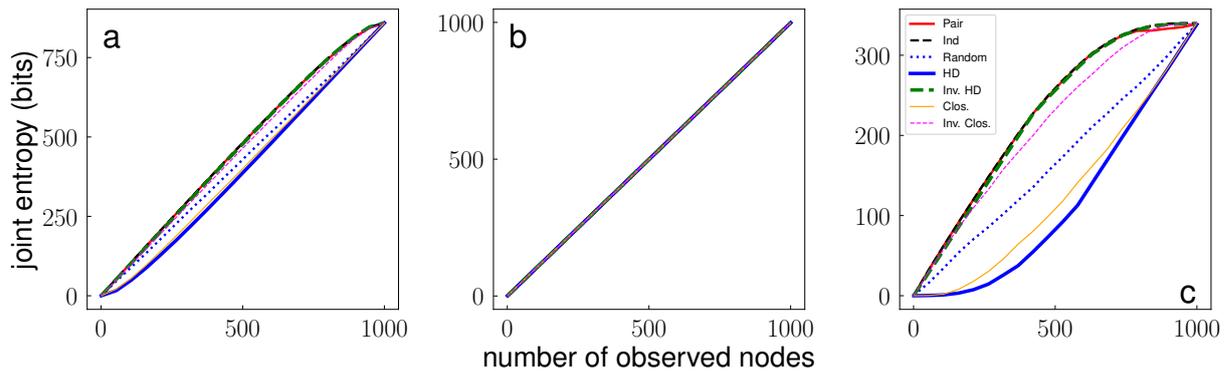

Figure SM42. MSIS model applied to the UCM. Panels a, b, and c correspond to the same parameters values as in panels b, c, and d of Fig. SM41, respectively. In all cases, joint entropy is measured with our approximation $\mathcal{H}_{pair}$. We consider different strategies $X$ to construct the observed set $\mathcal{O}_X$. In addition to the same already considered in Fig. SM41, we consider here also sampling strategies based on degree ($X$ = HD), and closeness centrality ($X$ = Clos.), where nodes are ranked according to those metrics. We consider also the inverse strategies, where nodes are ranked in increasing order based on the value of these metrics ($X$ = Inv. HD, and $X$ = Inv. Clos., respectively).

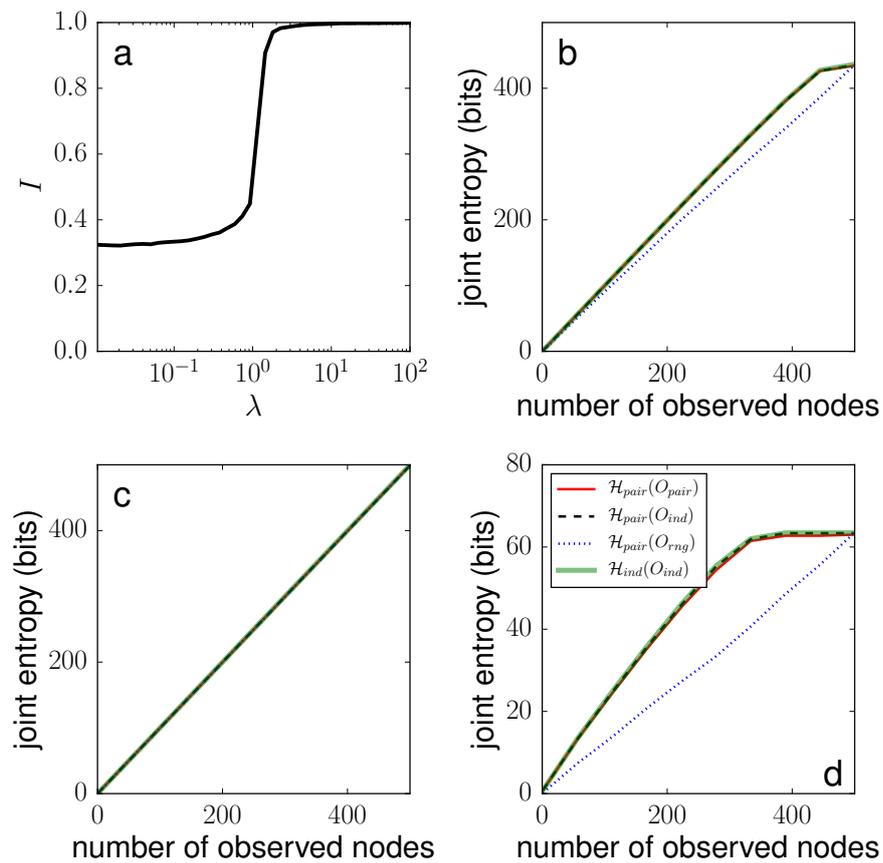

Figure SM43. We consider the MSIS model applied to the USATN. The description of panels a–d is the same as of Fig. SM41. The values of the infection rate $\lambda$ for panels b, c, and d are: b) $\lambda = 0.5$, c) $\lambda = 1.0$, d) $\lambda = 2.0$.

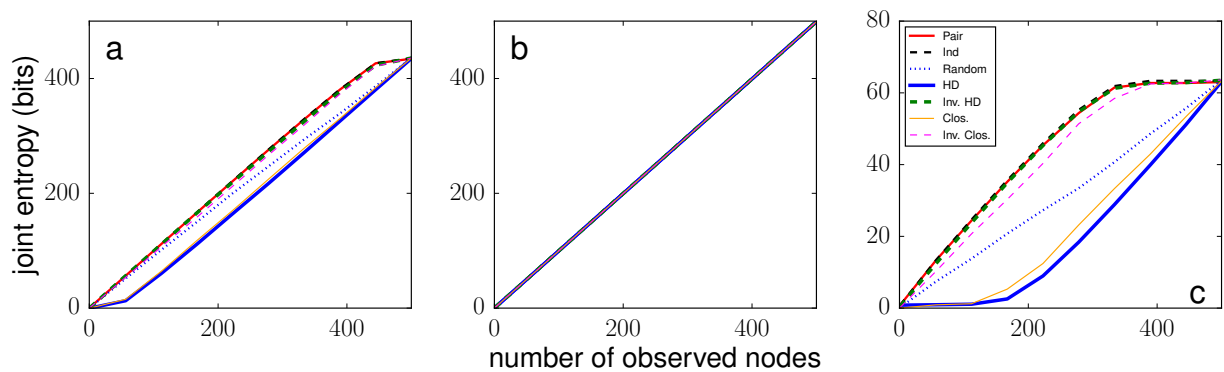

Figure SM44. MSIS model applied to the USATN. Panels a, b, and c correspond to the same parameters values as in panels b, c, and d of Fig. SM43, respectively. The description of the figure is ideintical to the one of Fig. SM42.

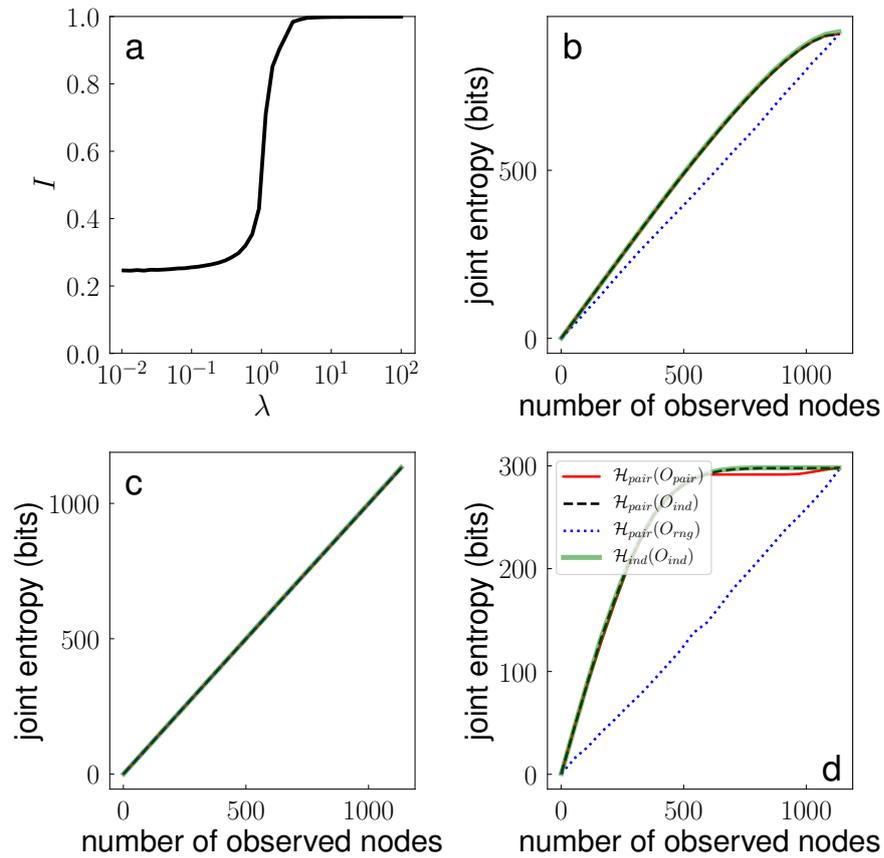

Figure SM45. We consider the MSIS model applied to the ECN. The description of panels a–d is the same as of Fig. SM41. The values of the infection rate $\lambda$ for panels b, c, and d are: b) $\lambda = 0.5$, c) $\lambda = 1.0$, d) $\lambda = 2.0$.

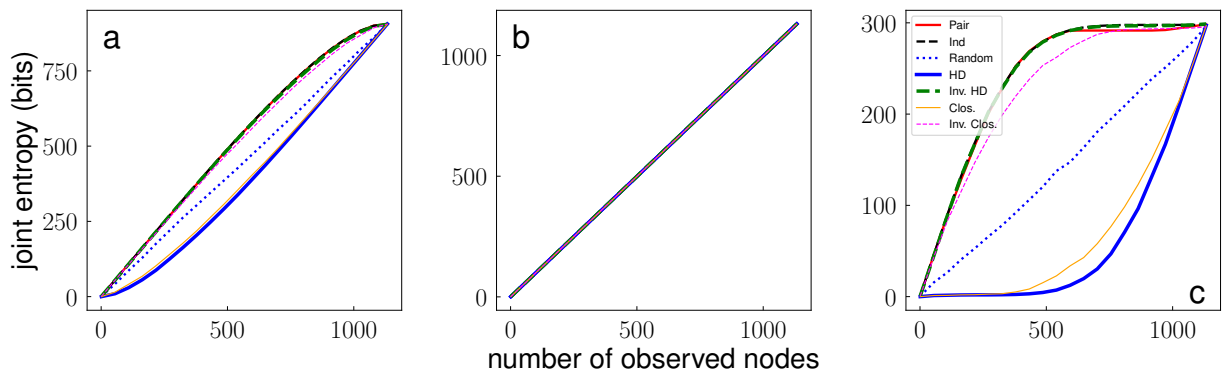

Figure SM46. MSIS model applied to the ECN. Panels a, b, and c correspond to the same parameters values as in panels b, c, and d of Fig. SM45, respectively. The description of the figure is ideintical to the one of Fig. SM42.

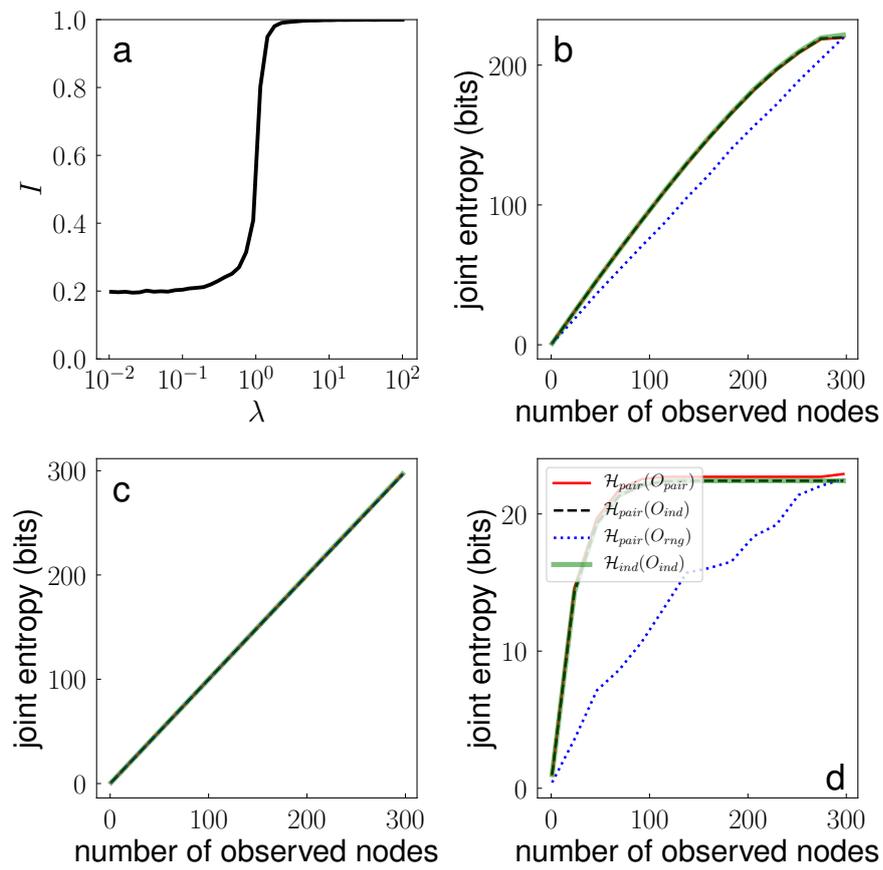

Figure SM47. We consider the MSIS model applied to the CENN. The description of panels a–d is the same as of Fig. SM41. The values of the infection rate $\lambda$ for panels b, c, and d are: b) $\lambda = 0.5$, c) $\lambda = 1.0$, d) $\lambda = 2.0$.

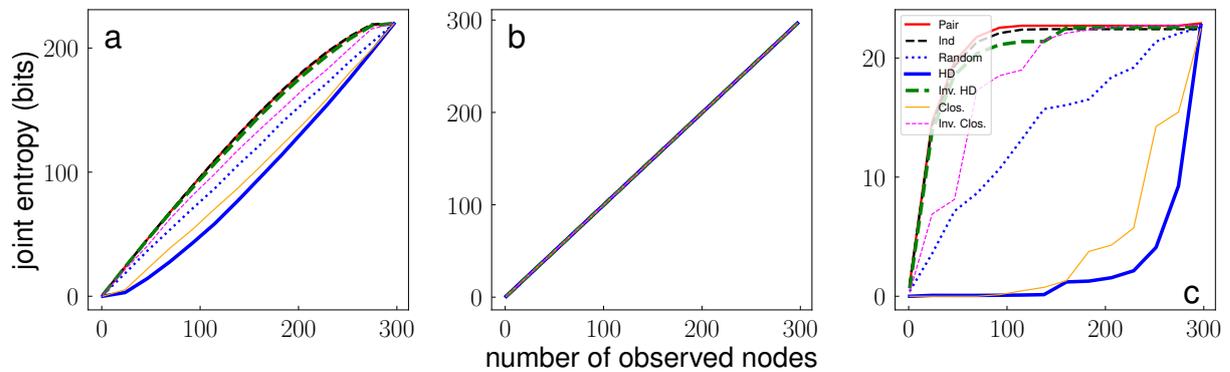

Figure SM48. MSIS model applied to the CENN. Panels a, b, and c correspond to the same parameters values as in panels b, c, and d of Fig. SM47, respectively. The description of the figure is ideintical to the one of Fig. SM42.

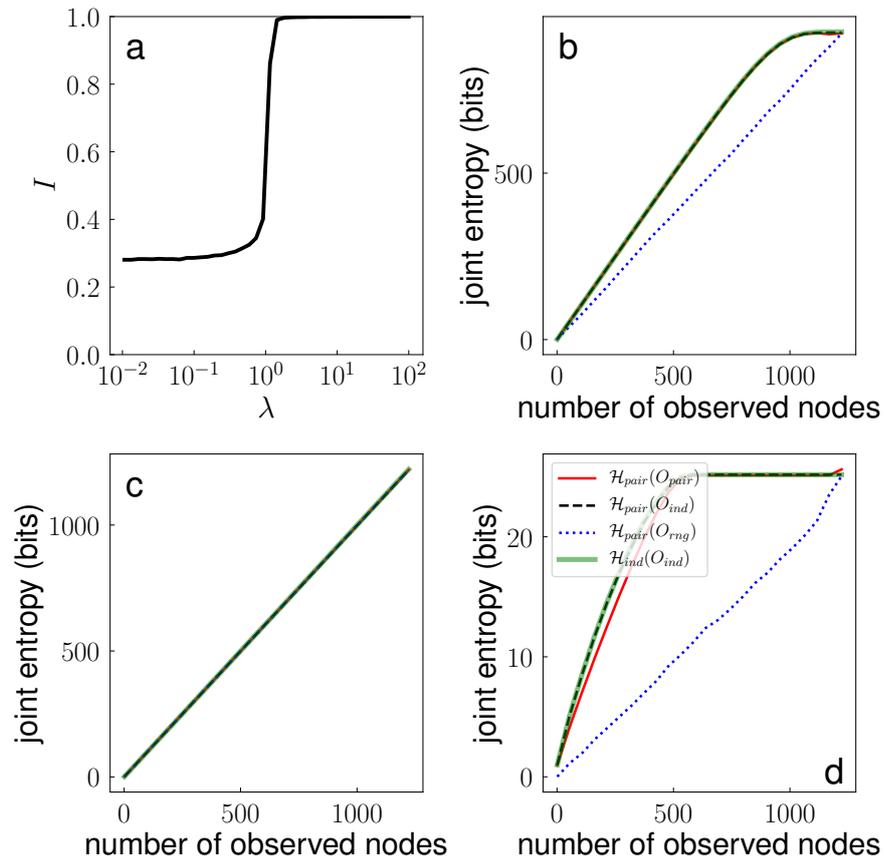

Figure SM49. We consider the MSIS model applied to the PBN. The description of panels a–d is the same as of Fig. SM41. The values of the infection rate $\lambda$ for panels b, c, and d are: b) $\lambda = 0.5$, c) $\lambda = 1.0$, d) $\lambda = 2.0$.

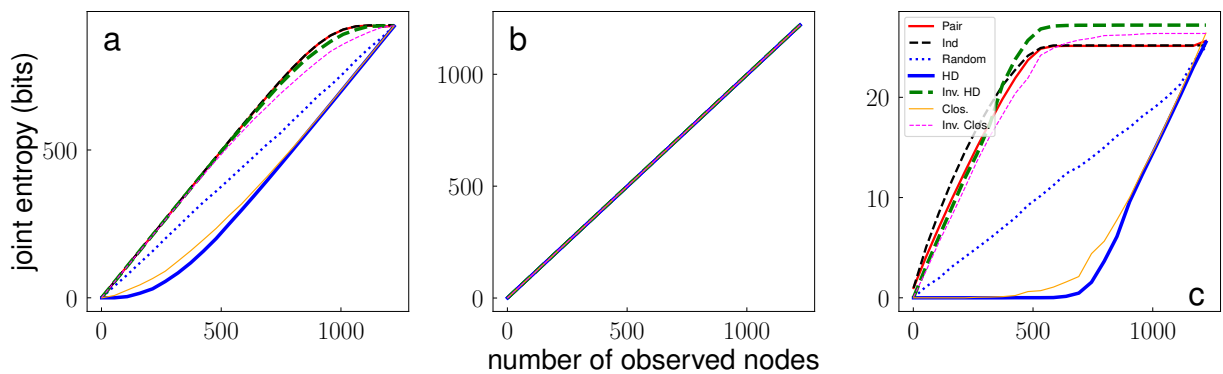

Figure SM50. MSIS model applied to the PBN. Panels a, b, and c correspond to the same parameters values as in panels b, c, and d of Fig. SM49, respectively. The description of the figure is ideintical to the one of Fig. SM42.

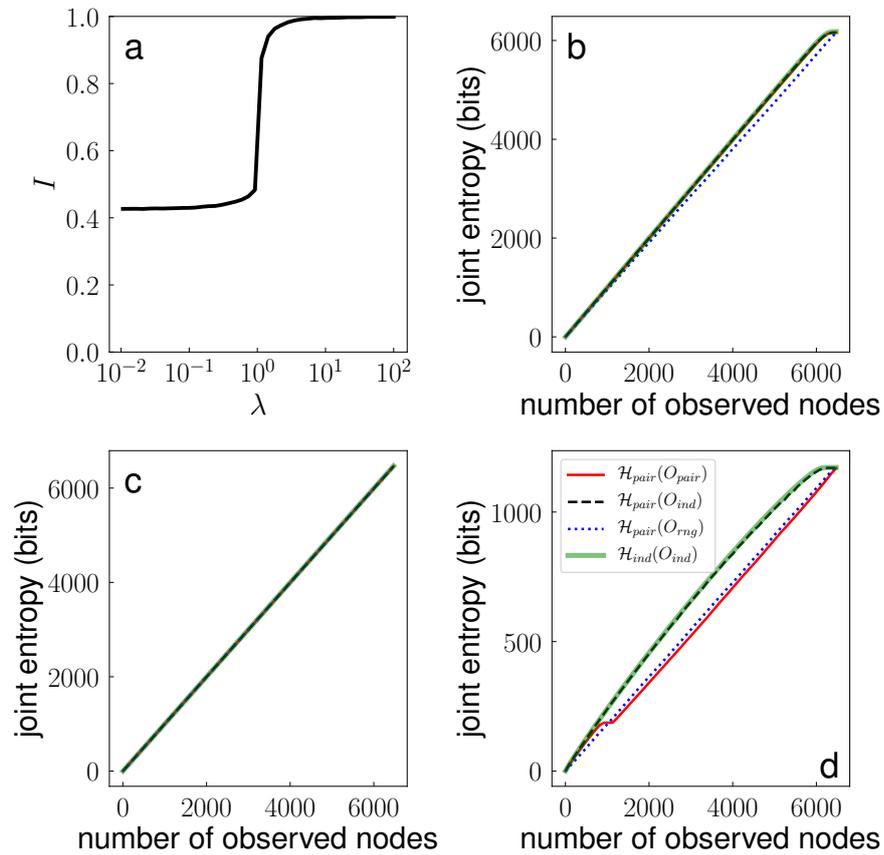

Figure SM51. We consider the MSIS model applied to the ASN. The description of panels a–d is the same as of Fig. SM41. The values of the infection rate $\lambda$ for panels b, c, and d are: b) $\lambda = 0.5$, c) $\lambda = 1.0$, d) $\lambda = 2.0$.

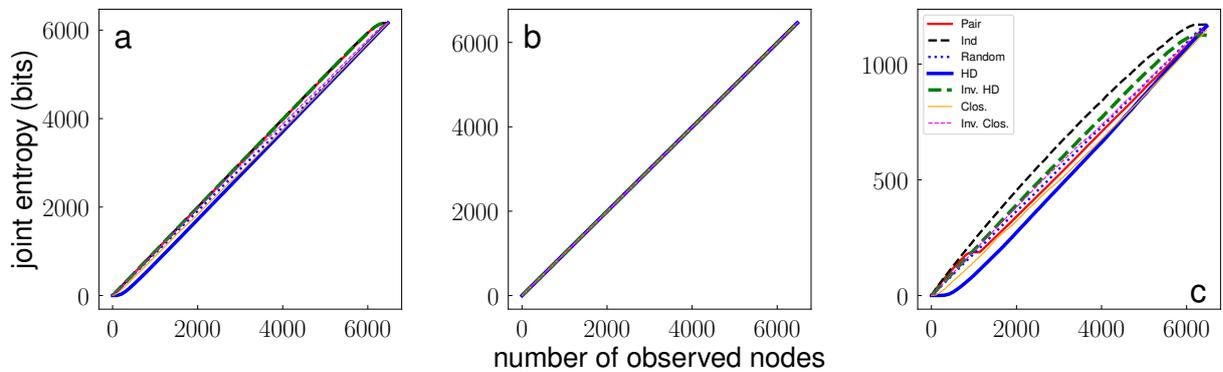

Figure SM52. MSIS model applied to the ASN. Panels a, b, and c correspond to the same parameters values as in panels b, c, and d of Fig. SM51, respectively. The description of the figure is ideintical to the one of Fig. SM42.

## Independent Cascade Model on a Star-like Network

To provide further evidence on how the details of a stochastic system may affect the identification of the best set of observed nodes, we study the Independent Cascade (IC) model on a star-like configuration. The topology is given by a network composed of $N$ peripheral nodes, with labels $p = 1, \ldots, N$, attached to a central node with label $c = N + 1$. In the IC model, nodes can be in three different states: $S$ for susceptible, $I$ for infected, and $R$ for recovered. The rules of the dynamics are simple. At every istant of time, every infected node passes the infection to its all neighbors that are in the $S$ state with probability $\lambda$. After that, the nodes recover, and they do longer participate in the dynamics. The process continues if new nodes have been infected, otherwise, it stops leaving us with final configuration where nodes are only in the states $S$ or $R$. Here, we focus our attention on the final configuration of the system. The infection probability $\lambda$ is one ingredient which we can play with. The second ingredient is the initial configuration of the system, namely $\mathbf{x}^{(0)}$, where we assume that for a generic node $i$ we can have $x_i^{(0)} = I$ or $S$. In the various cases below, we compute the entropy associated with the central node or with a peripheral node. The node with larger entropy is the one initially observed according to the maximum entropy principle.

### Unknown initial configuration

Let us consider the case in which we don't know anything about the initial configuration, so that every initial configuration has the same probability of appearance. The total number of possible initial configurations is $2^{N+1}$, and their associated probability is $2^{-(N+1)}$.

Suppose the total number of initially infected nodes is $n$. The probability that one these configurations includes among the selected nodes the central node is given by

$$p(x_c^{(0)} = I | N_I = n) = \frac{\binom{N}{n-1}}{\binom{N+1}{n}} = \frac{N!}{(n-1)!(N-n+1)!} \frac{n!(N+1-n)!}{(N+1)!} = \frac{n}{N+1} \ .$$

The probability that the total number of nodes that are initially infected is $n$ reads as

$$p(N_I = n) = 2^{-(N+1)} \binom{N+1}{n} . \tag{SM7}$$

The probability that the final state of the central node is $R$ depends on the initial configuration. In particular, this depends on whether the central node is initially infected or not. We have

$$p(x_c = R | N_I = n, x_c^{(0)}) = \begin{cases} 1 & , \text{if } x_c^{(0)} = I \\ 1 - (1-\lambda)^n & , \text{if } x_c^{(0)} = S \end{cases}$$

For a peripheral node, we have

$$p(x_p = R | N_I = n, x^{(0)}) = \begin{cases} \frac{n-1}{N} + \lambda \left(1 - \frac{n-1}{N}\right) & , \text{if } x_c^{(0)} = I \\ \frac{n}{N} + \left(1 - \frac{n}{N}\right) \lambda (1 - (1-\lambda)^n) & , \text{if } x_c^{(0)} = S \end{cases}$$

We know that

$$p(x_c = R) = \sum_{n=0}^{N+1} p(N_I = n) \sum_{x_c^{(0)} = S, I} p(x_c^{(0)} | N_I = n) \ p(x_c = R | N_I = n, x_c^{(0)})$$

thus

$$p(x_c = R) = \frac{1}{2^{N+1}} \sum_{n=0}^{N+1} \binom{N+1}{n} \left\{ \frac{n}{N+1} + \left(1 - \frac{n}{N+1}\right) [1 - (1-\lambda)^n] \right\} \tag{SM8}$$

For a peripheral node, we have instead

$$p(x_p = R) = \sum_{n=0}^{N+1} p(N_I = n) \sum_{x_c^{(0)} = S, I} p(x_c^{(0)} | N_I = n) \ p(x_p = R | N_I = n, x_c^{(0)})$$

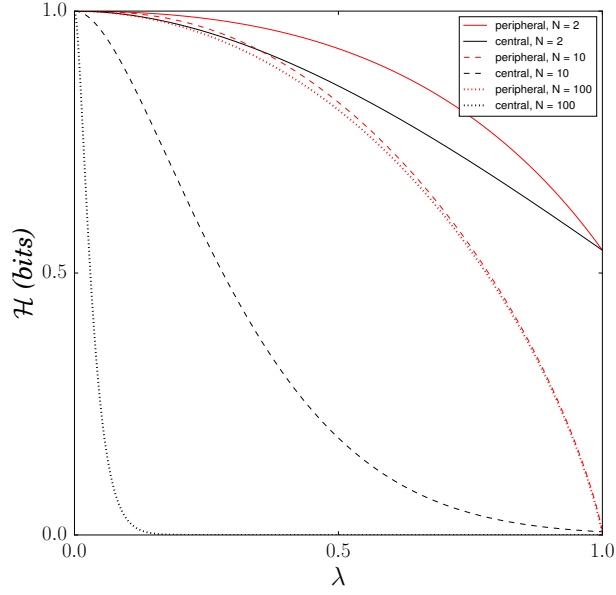

Figure SM53. Entropy derived from Eqs. (SM8) and (SM9).

thus

$$p(x_p = R) = \frac{1}{2^{N+1}} \sum_{n=0}^{N+1} \binom{N+1}{n} \left\{ \frac{n}{N+1} \frac{n-1}{N} + \frac{n}{N+1} \lambda \left( 1 - \frac{n-1}{N} \right) \right.$$
$$\left. \left( 1 - \frac{n}{N+1} \right) \frac{n}{N} + \left( 1 - \frac{n}{N+1} \right) \left( 1 - \frac{n}{N} \right) \lambda [1 - (1 - \lambda)^n] \right\}$$
(SM9)

Eqs. (SM8) and (SM9) can be finally used to compute the entropy associated with the central node or a generic peripheral node for given values of $\lambda$ and $N$ according to

$$\mathcal{H}(i) = p(x_i = R) \log_2[p(x_i = R)] + [1 - p(x_i = R)] \log_2[1 - p(x_i = R)] , \quad \text{with } i = c \text{ or } p .$$
(SM10)

*Initial configuration with a fixed density of infected nodes*

If we replace the probability of Eq. (SM7) with the binomial distribution

$$p(N_I = n) = \binom{N+1}{n} \rho^n (1 - \rho)^{N+1-n} ,$$
(SM11)

every single node has a probability $\rho$ to be initially in the $I$ state. We can therefore replace Eq. (SM8) with

$$p(x_c = R) = \sum_{n=0}^{N+1} \binom{N+1}{n} \rho^n (1 - \rho)^{N+1-n} \left\{ \frac{n}{N+1} + \left( 1 - \frac{n}{N+1} \right) [1 - (1 - \lambda)^n] \right\}$$
(SM12)

and Eq. (SM9) with

$$p(x_p = R) = \sum_{n=0}^{N+1} \binom{N+1}{n} \rho^n (1 - \rho)^{N+1-n} \left\{ \frac{n}{N+1} \frac{n-1}{N} + \frac{n}{N+1} \lambda \left( 1 - \frac{n-1}{N} \right) \right.$$
$$\left. \left( 1 - \frac{n}{N+1} \right) \frac{n}{N} + \left( 1 - \frac{n}{N+1} \right) \left( 1 - \frac{n}{N} \right) \lambda [1 - (1 - \lambda)^n] \right\}$$
(SM13)

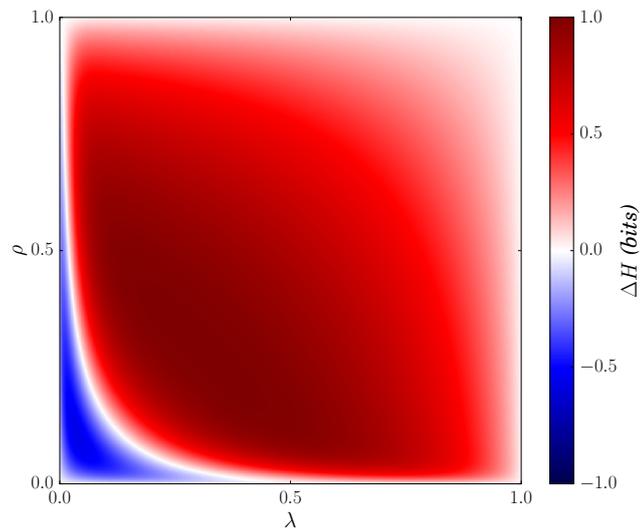

Figure SM54. Difference between the entropy computed with Eq. (SM13) and the one obtained with Eq. (SM12) as a function of the probabilities $\lambda$ and $\rho$. The size of the star is $N = 100$.

*Initial configuration with a single infected node*

Finally, let us focus only on starting configurations where the one and only one node is infected and all others are in the $S$ state. The probability that the central node is in the final configuration in state $R$ is

$$p(x_c = R) = \frac{1}{N+1} + (1 - \frac{1}{N+1})\lambda \qquad (SM14)$$

The probability that a generic peripheral node is in state $R$ in the final configuration is

$$p(x_p = R) = \frac{1}{N+1} + (1 - \frac{1}{N+1})\frac{1}{N}\lambda + (1 - \frac{1}{N+1})(1 - \frac{1}{N})\lambda^2 \qquad (SM15)$$

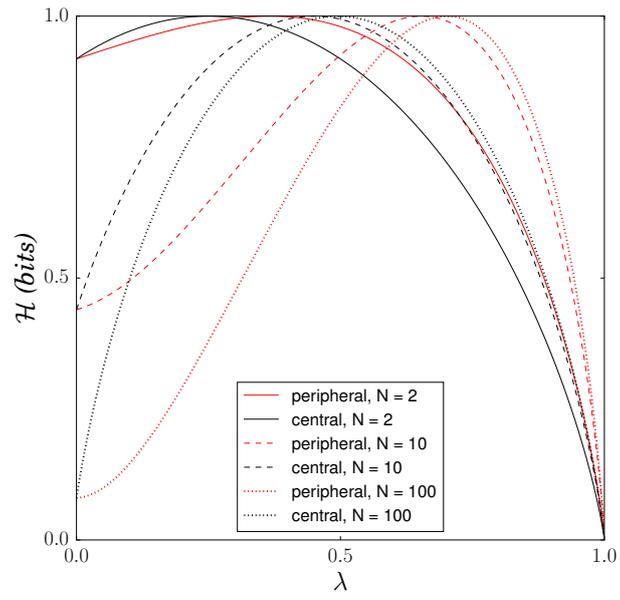

Figure SM55. Entropy derived from Eqs. (SM14) and (SM15).

––––––––––



\end{document}